\documentclass[12pt, preprint]{aastex}
\usepackage{epsfig}
\slugcomment{Draft version: \today}
\shorttitle{The Nonlinear Matter Power Spectrum}
\shortauthors{Z. Ma}

\begin{document}

\title{The Nonlinear Matter Power Spectrum
       \footnote{Presented as a dissertation to the Department of Astronomy 
                 and Astrophysics, The University of Chicago, in partial 
                 fulfillment of the requirements for the Ph.D. degree.}}
\author{Zhaoming Ma{\footnote{Email: mzm@oddjob.uchicago.edu}}}
\affil{University of Chicago, Department of Astronomy and Astrophysics
        \& Kavli Institute for Cosmological Physics, \\
        5640 South Ellis Ave., Chicago, IL 60637}

\begin{abstract}

   We modify the public PM code developed by Anatoly Klypin and Jon Holtzman
to simulate cosmology with arbitrary initial power spectrum and equation of
state of dark energy. With this tool in hand, we perform the following studies
on the matter power spectrum.

   With an artificial sharp peak at $k \sim 0.2 h{Mpc}^{-1}$ 
in the initial power spectrum, we find that 
the position of the peak is not shifted by nonlinear evolution. 
An upper limit of the shift at the level of $0.02\%$ is achieved by fitting
the power spectrum local to the peak using a power law plus a Gaussian.
This implies that, for any practical purpose, the baryon acoustic 
oscillation peaks in the matter power spectrum are not shifted by 
nonlinear evolution which would otherwise bias the cosmological distance
estimation. We also find that the existence of a peak in the linear power
spectrum would boost the nonlinear power at all scales evenly. This is 
contrary to what HKLM scaling relation predicts, but roughly consistent 
with that of halo model.

   We construct two dark energy models with the same linear power spectra
today but different linear growth histories. We demonstrate that their nonlinear
power spectra differ at the level of the maximum deviation of the 
corresponding linear power spectra in the past. 
Similarly, two constructed dark energy
models with the same growth histories result in consistent nonlinear power
spectra. This is hinting, not a proof, that linear power spectrum together
with linear growth history uniquely determine the nonlinear power spectrum.
Based on these results, we propose that linear growth history be included in
the next generation fitting formulas of nonlinear power spectrum.

   For simple dark energy models parametrized by $w_0$ and $w_a$, the existing
nonlinear power spectrum
fitting formulas, which are calibrated for $\Lambda$CDM model, work reasonably
well. The corrections needed are at percent level for the power spectrum and
$10\%$ level for the derivative of the power spectrum. 
We find that, for \cite{PD96} fitting formula, the corrections to the 
derivative of the power spectrum are independent 
of $w_a$ but changing with redshift. As a short term solution, a fitting form 
could be developed for $w_0$, $w_a$ models based on this fact.

\end{abstract}


\section{Introduction}

   For cosmologists, the clustering property of matter is a gold mine. 
Although we are at the beginning stage of
the operation, we have already been rewarded dearly: 
combined with WMAP \citep{Spergel03}, SDSS data constrains cosmological 
parameters to reasonably high precision (for example, the matter density 
of the universe 
$\Omega_m$ and the amplitude of the matter perturbation $\sigma_8$ to 
$\sim 10\%$ \citep{Tegmark04,Abazajian05,Seljak05}).
\citet{Fosalba03} and \citet{Scranton03} found direct evidence of
the mysterious dark energy through ISW effect using data from SDSS.
Also using SDSS data, the baryon acoustic oscillation (BAO) feature is used to
probe the expansion history of the universe and cosmological distances
\citep{Eisenstein05,Cole05}.
Weak lensing surveys \citep{Jarvis05} made the first attempt to constrain 
the time evolution of dark energy ... 
Moreover numerous ambitious future surveys of large scale structure are 
proposed, such as the Dark Energy Survey 
(DES\footnote{http://cosmology.astro.uiuc.edu/DES}),
PanSTARRS\footnote{http://pan-starrs.ifa.hawaii.edu}, 
Supernova/Acceleration Probe \citep[SNAP\footnote{http://snap.lbl.gov};][]
{SNAP} and 
Large Synoptic Survey Telescope (LSST\footnote{http://www.lsst.org}). 
From these surveys, we want to answer questions like: Is our
concordance cosmology model correct? What is the nature of the mysterious
dark matter and dark energy? Do we need to modify gravity theory? Is the 
universe flat or curved?

    On the other hand, we have tremendous success in theoretical modeling
of structure formation in our universe. Starting with a handful of cosmological
parameters, we can compute the linear theory matter power spectrum with 
high accuracy \citep[e.g.,\ ][]{Seljak03}.  However, we measure the nonlinear 
matter or galaxy power spectrum from large-scale structure surveys. Our
understanding of the connection between linear and nonlinear power spectrum
is improving rapidly along with our ability to simulate the complex processes
involved.

\subsection{Baryon acoustic oscillations}

    The baryon acoustic oscillation (BAO) features in the late time matter 
clustering, characterized by a single peak in the correlation
function and oscillations in the matter power spectrum, 
are imprints of the acoustic oscillations in the early universe
which produces the peaks and troughs in the CMB angular power spectrum 
\citep{Peebles70,Sunyaev70}. 
Although the BAO features in the matter power spectrum are much weaker compared
to those of the CMB because the dominating dark matter did not participate in
the acoustic oscillations, these features can be measured with high
accuracy with sufficiently large surveys. Indeed, the first detection was made
in the SDSS survey in 2005 \citep{Eisenstein05,Cole05}.

With the large galaxy surveys proposed, e.g. DES, PANSTARRS, SNAP, and LSST, 
BAO is emerging as an important method to measure the expansion history of 
the universe and cosmological distances
\citep{Eisenstein98, Cooray01, Blake03, Hu03, Seo03, Linder03, Matsubara04,
 Amendola05, Blake05, Glazebrook05, Dolney06}. 
The main theoretical uncertainties, including galaxy bias, nonlinear 
evolution and redshift space distortions, are investigated in detail by 
\cite{White05}, \cite{Eisenstein_Seo06_1}, \cite{Eisenstein_Seo06_2}
\cite{Huff06} and \cite{Smith06}.
Nonlinear evolution smears BAO peaks due to mode-mode coupling. This
effect makes BAO signal weaker and harder to detect. An even more worrying
effect is that nonlinear evolution could potentially shift the peaks which
would bias the distance measurement from which the cosmological parameters 
are inferred. 

\cite{Eisenstein_Seo06_1} made 
an order of magnitude estimation of the amount of shift of BAO peaks due
to nonlinear evolution.
They use spherical collapse model to estimate how much the overdensity within 
$150\,{\rm Mpc}$ has grown if the center were over-dense;
and the change in radius is a third of that in the over density.
Their conclusion is that the change in scale is on the order of $O(10^{-4})$. 
In this work, we use N-body simulations to test how much the peaks are
shifted due to nonlinear evolution.

\subsection{Fitting formulas of the nonlinear matter power spectrum}

   Calibrated by simulations, fitting formulas of the nonlinear 
power spectrum have been developed \citep{PD96,Smith03,McDonald05}. 
Their precisions, on the order of
$10-20\%$, are sufficient for most of the current applications. Future
surveys would have statistical uncertainties so low that
fitting formulas with improved accuracy are highly desired \citep{Huterer05}. 
As a step toward fulfilling this goal,
we test the assumptions that go into the afore mentioned
fitting formulas and propose the necessary ingredients to build a more
accurate one. First, let us briefly review the existing fitting formulas.

    In a series of papers \citep{Hamilton91, PD94, Jain95, PD96}, 
the first fitting formula of the nonlinear power spectrum is
worked out. The foundation of the fitting formula is the so called HKLM
relation which has two parts. 
The first part is to connect the nonlinear 
scale, $r_{NL}$, to the linear scale, $r_{L}$, by volume-averaged correlation 
function $\bar{\xi}(r)$,
\begin{equation}
  r_{L} = [ 1 + \bar{\xi}_{NL}(r_{NL}) ]^{1/3} r_{NL} \,.
\end{equation}
The reasoning behind this is pair conservation \citep{Peebles80}. The second
part of the HKLM procedure is to conjecture that the nonlinear correlation
function is a universal function of the linear one,
\begin{equation}
  \bar{\xi}_{NL}(r_{NL}) = f_{NL} [\bar{\xi}_{L}(r_{L})] \,.
\end{equation}
If one assumes that  $\bar{\xi}(r)$ is equivalent to matter power spectrum 
at an effective wavenumber $k$, then we have the power spectrum version of 
the HKLM relation,
\begin{equation}
  k_{L} = [ 1 + \Delta^2_{NL}(k_{NL}) ]^{1/3} k_{NL} \,,
\end{equation}
\begin{equation}
  \Delta^2_{NL}(k_{NL},t) = f_{NL} [\Delta^2_{L}(k_{L},t)] \,.
\end{equation}
It is commonly referred to as Peacock \& Dodds fitting formula and $f_{NL}$
is calibrated using N-body simulation assuming $\Lambda$CDM cosmology.

    Another approach of calculating the nonlinear power spectrum is to use
the halo model \citep[See][for a review]{Sheth02}, which splits up the contribution to the power 
spectrum into one halo and two halo terms,
\begin{equation}
  P_{NL}(k) = P^{1h}(k) + P^{2h}(k) \,.
\end{equation}
The one halo term $P^{1h}(k)$ is the contribution from two particles that
reside in the same halo,
\begin{equation}
  P^{1h}(k) = \int dm n(m) \left ( {m \over {\bar{\rho}}} \right )^2
              |u(k | m)|^2 \,,
\label{eqn:P1h}
\end{equation}
and the two halo term is the contribution from two particles that belong
to two different halos,
\begin{equation}
  P^{2h}(k) = \left [ \int dm n(m) \left ( {m \over {\bar{\rho}}} \right )
              b(m) | u(k | m) | \right ]^2 P_{L}(k) \,.
\label{eqn:P2h}
\end{equation}
Here $m$ is the mass of the halo, $\bar{\rho}$ is the mean density of the
universe, and $u(k | m)$ is the Fourier transform of the profile of a halo
\citep[NFW profile][for example]{NFW97} with mass $m$,  $n(m)$ is the 
halo mass function which has the following form 
\citep{Press74, Jenkins01, Sheth99}, 
\begin{equation}
   n(m) = {{\bar{\rho}} \over m} f(\nu) {{d \nu} \over {dm}} \,,
\end{equation}
where $\nu = \delta_c / \sigma(m) $, $\sigma(m) $ is the rms of the linear 
density field smoothed by a spherical top-hat filter $W(k,m)$,
\begin{equation}
   \sigma^2(m) = \int_0^\infty dlnk \Delta^2_L(k) W^2(k,m) \,.
\label{eqn:sigmaM}
\end{equation}
Note that adding a sharp spike in the matter power spectrum would only
modify $\sigma(m)$ below a mass threshold which is determined by the
location of the spike. The modification to the mass function $n(m)$ 
would also only occur below the same mass threshold. This could be
easily shown by adding a $\delta$-function to $\Delta^2_L(k)$ in
equation\,\ref{eqn:sigmaM}.

    Finally, based on a fusion of the halo model and an HKLM scaling, 
\cite{Smith03} provided an empirical fitting formula with improved 
performance (rms error better than $7\%$ is claimed). In detail, they
replaced the two halo term with a \cite{PD96} type fitting form with an
exponential cut off at nonlinear $k$ to deal with translinear regime
better; and a fitting form to the one halo term. Again, the fitting
forms are calibrated using $\Lambda$CDM simulation.

    To summarize, fitting formulas of the nonlinear matter power spectrum
have been developed for  $\Lambda$CDM cosmology with precisions on the
order of $10-20\%$. The foundations for these fitting formulas are the
HKLM scaling relation and halo model. A less obvious assumption behind
all these fitting formulas is that the nonlinear power spectrum is
determined by the linear power spectrum at the same epoch, i.e. the
history of the linear power spectrum does not affect the nonlinear one.

    In order to achieve the $1\%$ accuracy required by future surveys
\citep{Huterer05}, 
and with more general parametrization of dark energy models, 
what approach shall we take to build new 
fitting formula of the nonlinear power spectrum? To address the issue
we will test the assumptions that go into the existing fitting formulas 
using carefully constructed numerical experiments. In the end we will decide
which assumptions should be abandoned, kept, and further included. 

    Although all the existing fitting formulas are calibrated for 
$\Lambda$CDM cosmology, it has been a common practice to apply them to
dark energy models in general. We will quantify how 
well this procedure works.

    The rest of the paper is organized as follows: 
In section 2 we briefly introduce the PM simulations.
We demonstrate the behavior of the peak in the initial power spectrum in 
section 3, and discuss what we could learn from it.
We test basic building blocks of the fitting formulas in section 4.
In section 5 we evaluate the performance of the fitting formulas when 
applied to general dark energy models. 
Finally, we summarize our findings.

\section{The PM simulation}

    The public PM code developed by Anatoly Klypin and Jon Holtzman is used 
for this work. The source code and manual are available at 
\newline \hspace*{5mm} http://charon.nmsu.edu/${\sim}$aklypin/PM/pmcode/
\newline \hspace*{16mm} pmcode.html
\newline
The code assumes $\Lambda$CDM cosmology. In order to simulate cosmology with 
an arbitrary equation of state of dark energy, $w(z)$, or/and arbitrary 
initial power spectrum, we have modified the code appropriately.
In this section, we lay out the basic formalism used in the simulation and 
the tests we have performed.

\subsection{Formalism}

  We will use $r$, $t$ and $v$ as position, time and velocity variables in 
physical space and $x$, $\tau$ and $u$ as the corresponding variables in 
comoving space.

  In physical space the Poisson equation is,
\begin{equation}
 \nabla_r^2 \Phi = 4 \pi G \rho \,,
\end{equation}
where $\Phi$ is the gravitational potential, $G$ is the gravitational constant,
and $\rho$ is the density of the universe including all constituencies.
  If we define a new potential $\phi$ as,
\begin{equation}
 \phi \equiv \Phi + {1 \over 2} { {\ddot a} \over a } x^2 \,,
\end{equation}
where over-dot represents derivative with respect to $\tau$ and
$a$ is the scale factor,
then, in comoving coordinates, the Poisson equation becomes,
\begin{equation}
  \nabla_x^2 \phi = 4 \pi G \rho a^2 + 3 { {\ddot a} \over a } \,.  
\end{equation}
Using Fredmann equation to substitute the ${ {\ddot a} / a }$ term,
we have,
\begin{equation}
  \nabla_x^2 \phi = 4 \pi G a^2 (\rho - {\bar{\rho}}) 
                  \equiv 4 \pi G a^2 \delta \rho \,,
\label{eqn:poisson}
\end{equation}
here ${\bar{\rho}}$ is the mean density of the universe. The corresponding
equations of motion in comoving space are,
\begin{equation}
  p \equiv m u = m \dot{x} \,; \,\,\, \dot{p}=m \dot{u} = -m \nabla_x \phi \,,
\label{eqn:eom}
\end{equation}
with $m$ the mass of the particle.

    The PM code solves the Poisson equation \ref{eqn:poisson} and the equations
of motion \ref{eqn:eom}. It is convenient to make the variables dimensionless
by choosing suitable units. Anatoly Klypin's PM code defines the code variables
(denoted by $\sim$) as,
\begin{equation}
  x = \tilde{x} x_0 \,,\,\,\,  
  p/m = {\tilde{p} \over a} p_0 \,,\,\,\,
  \phi = \tilde{\phi} \phi_0 \,,\,\,\,  
  \rho = {\tilde{\rho} \over {a^3}} \rho_0 \,,
\end{equation}
with the units,
\begin{displaymath}
  {x_0} = {L_{\rm box} / {N_{\rm g}^{1/3}}} \,,\,\,\,  
  {p_0} = x_0 H_0 \,,
\end{displaymath}
\begin{equation}
  {\phi_0} = (x_0 H_0)^2 \,,\,\,\,  
  {\rho_0} = {3H_0^2 \over {8 \pi G}} \Omega_m \,.
\end{equation}

Equation\,\ref{eqn:poisson} and \ref{eqn:eom}, in terms of these dimensionless
variables, become,
\begin{equation}
 {\tilde{\nabla}}^2 \tilde{\phi} = {3 \over 2} {\Omega_m \over a} ( \tilde{\rho} - 1),
\label{eqn:poissonCode}
\end{equation}
\begin{equation}
 {d \tilde{p} \over {da}} = -F(a) {\tilde{\nabla}} \tilde{\phi} \,,\,\,\,
 {d \tilde{x} \over {da}} = F(a) {\tilde{p} \over {a^2}} \,,
\label{eqn:eomCode}
\end{equation}
where
\begin{eqnarray}
 F(a) &=& {H_0 / {\dot{a}}} \\
     &=& {\sqrt{a} \over {\sqrt{\Omega_m + \Omega_{\Lambda} a^3 
     \exp \left [-3 \int_{a=1}^a (1 + w(a')) dlna'\right ] }} } \, .
       \nonumber
\end{eqnarray}
Here $w(a)$ is the equation of state of dark energy which has been generalized
to an arbitrary function of $a$.

\subsection{Brief summary of the PM code}

   PM code solves equation\,\ref{eqn:poissonCode} and \ref{eqn:eomCode} in
four main steps.
\begin{itemize}
   \item{setup the initial positions and velocities of the dark matter 
         particles}
   \item{solve the Poisson equation using the density field estimated with
         current particle positions}
   \item{Advance velocities using the new potential}
   \item{Update particle positions using new velocities}
\end{itemize}

The initial condition is setup using Zel'dovich approximation 
\citep{Zeldovich70, Crocce06}.
The density assignment is done using Cloud-in-Cell (CIC) interpolation, which 
is the method that linearly interpolates particle mass to its eight 
neighboring cells. The Poisson
equation is solved using the Fast Fourier Transform (FFT) and the particle 
positions and velocities are solved using the leapfrog integration scheme.
For further details of the implementation of these schemes, please check
out the excellent write-up by Andrey Kravtsov at
\newline \indent
   http://astro.uchicago.edu/$\sim$andrey/Talks/PM/pm.pdf \,.
\newline

\subsection{Code testing}

    We setup a few test runs to determine the parameters to use in the 
simulations. These include the number of particles $N_{\rm p}$, number of grid 
cells $N_{\rm g}$, the starting redshift $z_{\rm ini}$, evolving time step $da$ 
and the box size $L_{\rm box}$. 

    Ideally, the larger $N_{\rm p}$ and $N_{\rm g}$ are the better. In practice,
they are limited by the computation resources. For a standard run used
in this work, $N_{\rm p}=256^3$ and $N_{\rm g}=512^3$, one time step including the power
spectrum evaluation takes about 2.5\,min to 6\,min on an easily accessible
Pentium box. Approximately 500Mb of memory is required. For a bigger
run with $N_{\rm p}=512^3$ and $N_{\rm g}=1024^3$, which is mainly used to check the 
uncertainties of the standard runs, one step takes about 8 times longer
and memory usage is on the level of 4Gb. 

\begin{figure}[ht]
\begin{center}
\includegraphics[width=3.4in]{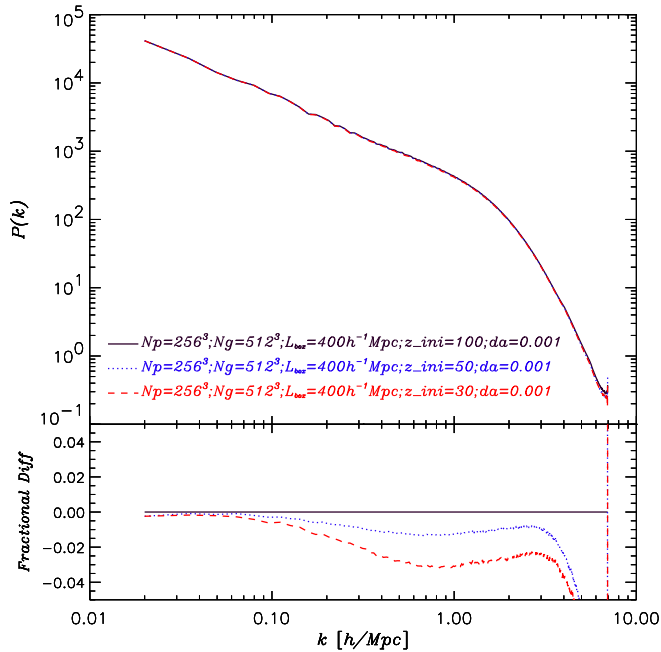}
\end{center}
\caption {The effect of initial redshift of the simulation on the
                matter power spectrum at $z=0$. The initial redshift $z_{\rm ini}$
                and the step size $da$ are listed in the legend.
                The Nyquist frequency is $k_{\rm Ny} = 4 \,h {\rm Mpc}^{-1}$.
                All of our simulations use $z_{\rm ini}=50$ unless stated
                otherwise.
\label{fig:zini100_50_30da0.001}}
\end{figure}

\begin{figure}[ht]
\begin{center}
\includegraphics[width=3.4in]{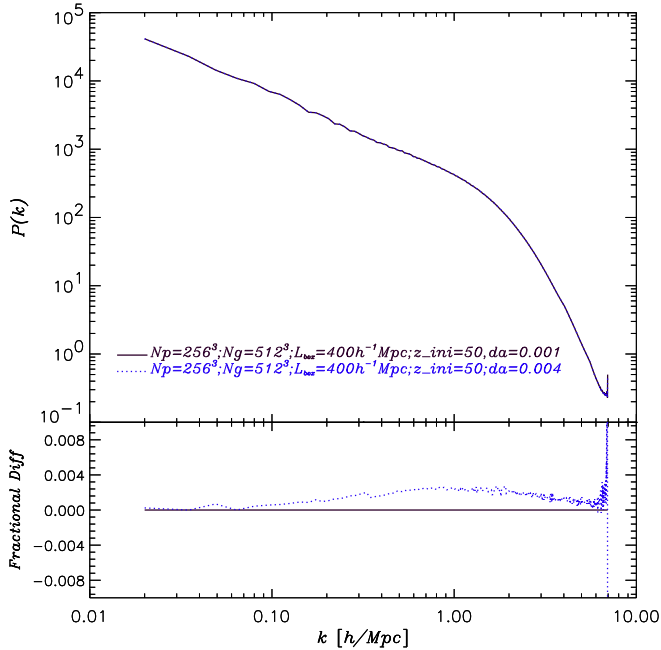}
\end{center}
\caption[Step size da: 0.001 vs 0.004 (zini=50)]
               {Effect of the step size of the simulation on the matter
                power spectrum at $z=0$. The initial redshift $z_{\rm ini}=50$
                and the stepsizes $da$ are listed in the legend.
                The Nyquist frequency is $k_{\rm Ny} = 4 \,h {\rm Mpc}^{-1}$.
                All of our simulations use $da=0.004$ unless stated
                otherwise.
\label{fig:zini50_50da0.001_4}}
\end{figure}

    Figure\,\ref{fig:zini100_50_30da0.001} shows the differences of
the matter power spectrum at $z=0$ due to different initial redshift
of the simulation, $z_{\rm ini}$. 
Simulations with lower $z_{\rm ini}$ under 
estimate the nonlinear power spectrum. Relative to $z_{\rm ini}=100$,
$z_{\rm ini}=50$ and $z_{\rm ini}=30$ have power deficits of $1.5\%$ and 
$3\%$ respectively. These are in good agreement with the results from
\cite{Crocce06}. For most of our applications, we use the ratio
of the power spectrum. Percent level power deficit is not a concern
for us. So we choose $z_{\rm ini}=50$ as our standard initial redshift 
unless stated otherwise. As to the step size,
figure\,\ref{fig:zini50_50da0.001_4} shows that, with $z_{\rm ini}=50$,
the power spectrum at $z=0$ differs by less than $0.3\%$ between $da=0.004$ 
and 0.001. Choosing $da=0.004$ as our standard step size, a full standard
run takes anywhere between 10 hours and a day.

\begin{figure}[ht]
\begin{center}
\includegraphics[width=3.4in]{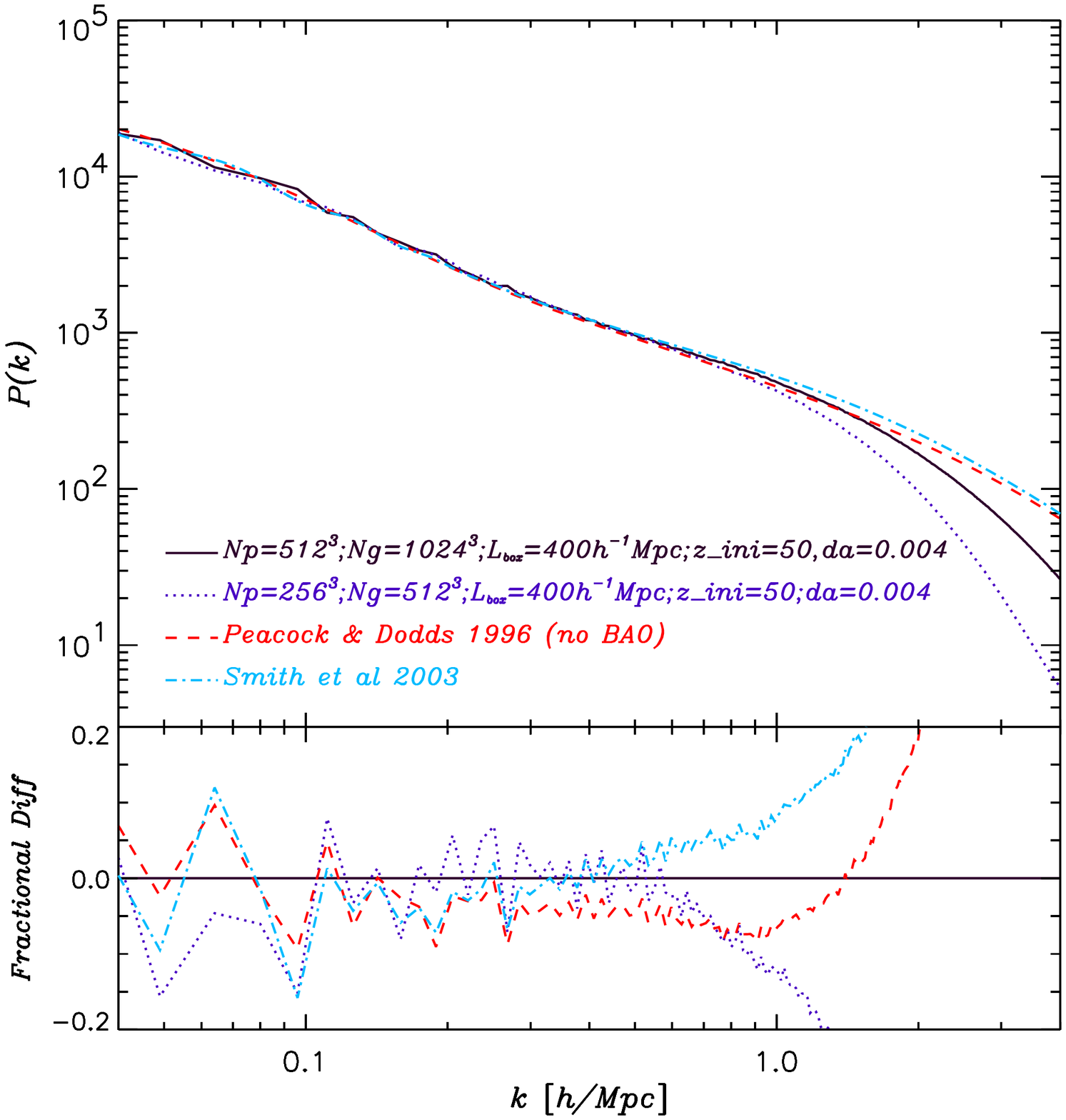}
\end{center}
\caption[Comparing w=-1 power spectrum (Lbox=400)] {Comparison of
                    the matter power spectrum, at $z=0$, from the
                    simulation with these from fitting formulas of
                    \cite{Smith03} and \cite{PD96}.
                    The upper panel shows the matter power spectra $P(k)$ and the
                    lower panel shows the fractional difference of $P(k)$
                    between the higher resolution simulation with $N_{\rm p}=512^3$
                    and the rest.  The simulation
                    box is 400 $h^{-1} {\rm Mpc}$ and the corresponding
                    Nyquist frequencies are $k_{\rm Ny} = 4$ and $8 \,h {\rm Mpc}^{-1}$
                    for simulations with $N_{\rm p}=256^3$ and $N_{\rm p}=512^3$ 
                    respectively.
                    The initial power spectrum without BAO peaks is used for
                    the Peacock \& Dodds fitting formula.
\label{fig:PkAllLbox400}}
\end{figure}

    The box size of the simulation is more determined by the $k$ range
that a particular application requires than anything else. 
The Nyquist frequency $k_{\rm Ny} \equiv N^{1/3}_g \pi / L_{\rm box}$ is the 
highest mode that is resolved by the simulation. The lowest mode is
$ 2 \pi / L_{\rm box}$. Because of aliasing, not all the $k$ modes in this
range are faithfully calculated. To determine the precision of the power
spectrum, we use these from simulations with higher resolution 
($N_{\rm p}=512^3$ and $N_{\rm g} = 1024^3$) 
with which we compare the results from our standard simulation
($N_{\rm p}=256^3$ and $N_{\rm g} = 512^3$).
With $L_{\rm box} = 400 h^{-1}{\rm Mpc}$, figure\,\ref{fig:PkAllLbox400}
shows that our standard run agrees with the higher resolution run perfectly
for $k < 0.6 h {\rm Mpc}^{-1}$ and the agreement worsens to about $10\%$ at
$k = 1.0 h {\rm Mpc}^{-1}$.

\begin{figure}[ht]
\begin{center}
\includegraphics[width=3.4in]{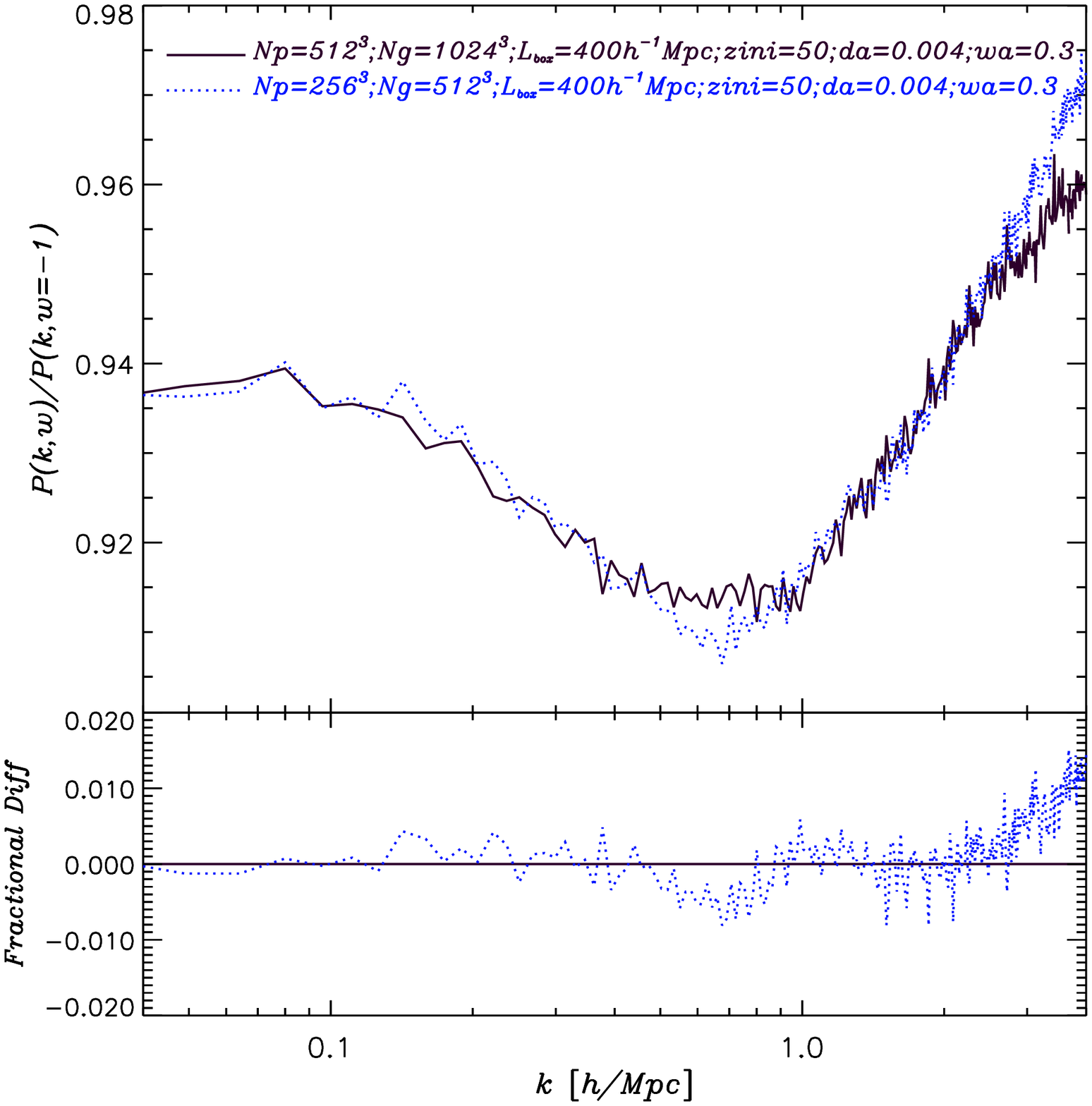}
\end{center}
\caption[Ratio of power spectra] {The ratio of
                    the matter power spectra of $w_a=0.3$ model to
                    that of $\Lambda$CDM model at $z=0$.
                    Different curves are from
                    simulations with different resolutions which are
                    labeled in the legend. The
                    Nyquist frequencies are $k_{\rm Ny} = 4$ and $8 \,h {\rm Mpc}^{-1}$
                    for simulations with $N_{\rm p}=256^3$ and $N_{\rm p}=512^3$ 
                    respectively.
\label{fig:Pk_ratio}}
\end{figure}

     For all of our applications, except for the test of the shifting of peaks 
in the power spectrum, we use the ratio of power spectra instead of 
the absolute power. This practice takes
care of the sample variance.
Figure\,\ref{fig:Pk_ratio} shows the uncertainty levels of the ratio.
Again, we compare our standard run with the one that has higher resolution
($N_{\rm p}=512^3$ and $N_{\rm g} = 1024^3$). For each simulation setup, we run
two cosmological models and take the ratio of their power spectra. 
One cosmological model is $\Lambda$CDM 
($w_0 \equiv w(z=0) =-1$ and $w_a \equiv -{{dw} / {da}}|_{a=1}=0$) 
and the other is $w_0 = -1$ and $w_a=0.3$ model. 
According to figure\,\ref{fig:Pk_ratio}, the
ratios of the power spectra in simulations of different resolutions 
agree to better than $0.5\%$ all the way to $k \sim 3 \,h{\rm Mpc}^{-1}$.

\section{the Fate of a Peak in the Initial Power Spectrum}

\subsection{Setting up the experiment}

    To address the issue of whether nonlinear evolution of the power spectrum
would shift the position of BAO peaks, we construct the following numerical
experiment.
We put an artificial sharp peak in the otherwise smooth initial matter power
spectrum and evolve it using a PM simulation (\cite{Shandarin90,Bagla97} used 
similar setups to test the transfer of power among different scales). 
The initial peak is
located at $k \sim 0.2 h{\rm Mpc}^{-1}$ with a $FWHM = 0.013 h{\rm Mpc}^{-1}$ and the 
maximum at $k=0.2051 h{\rm Mpc}^{-1}$. At the maximum, $P(k)$ is ten times what it
would be without the peak. The location of the peak is at about where the
third peak of the BAO sits and deep in the nonlinear regime today. The reason
to use a sharp peak is to make it easier to detect peak shifting without
using any sophisticated analysis. To set the peak amplitude high is to overcome the
sample variance of the simulation which is at least at the level of the
BAO peaks ($10\%$). Techniques have been developed to get around sample
variance and make BAO features easily seen in simulations. For example,
the {\it Millennium Simulation} group \citep{Springel05} corrects the power 
spectrum of the actual realization of the initial fluctuations in their 
simulation to the expected input power and apply these scaling factors 
at all other times. It is not clear that these scaling factors are not
affected by nonlinear evolution.
To test the physics behind it, our approach is much cleaner.

   The requirement for the simulation is such that $k=0.2 h{\rm Mpc}^{-1}$ is 
well resolved and the peak is sampled with at least a few points in $k$ space.
With these considerations, we use a simulation with box size of $2 h^{-1}{\rm Gpc}$,
number of particles $N_{\rm p} = 512^3$ and number of grid cells $N_{\rm g} = 1024^3$. The
corresponding Nyquist frequency is $1.6 h {\rm Mpc}^{-1}$ and the peak is in 
the well resolved scale. With this simulation setup, the peak is sampled with
five points in $k$ space.

\subsection{The shifting of the peak}

\begin{figure}[ht]
\begin{center}
\includegraphics[width=3.4in]{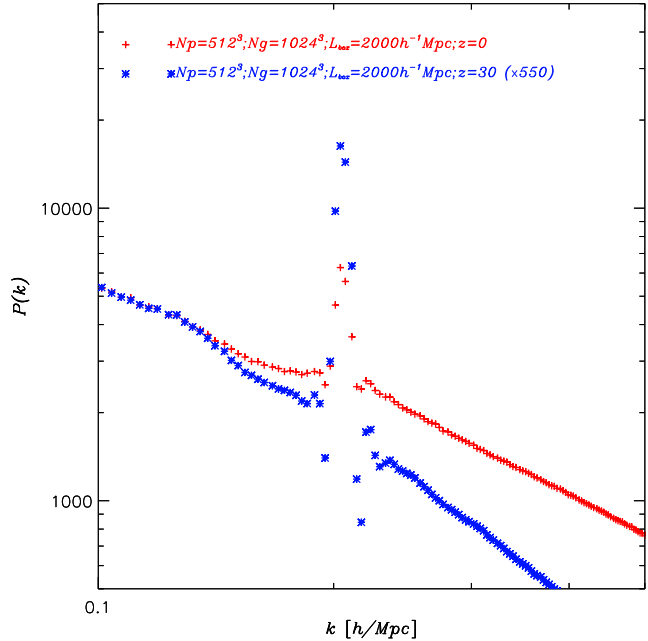}
\end{center}
\caption[Peak shift] {Testing the shift of peaks in the initial power spectrum.
                      The blue asterisks show the linear power spectrum at
                      $z=0$. The nonlinear power spectrum at $z=0$ is shown
                      using red plus signs. While the peak is being smeared
                      by nonlinear evolution, the position of the peak is not
                      shifted at all.
\label{fig:peakShift}}
\end{figure}

    Figure\,\ref{fig:peakShift} shows the position of the peak at $z=30$ 
and $z=0$.  Table\,\ref{tbl:peakNum}
lists the positions $k$ and the corresponding $P(k)$ for the points within 
the peak.  
\begin{deluxetable}{|c|c|c|c|c|c|}

\tablecolumns{6}
\tablewidth{0pt}
\tableheadfrac{0}
\tablecaption{Peak position and height.
\label{tbl:peakNum}}
\tablehead{
\colhead{ } & \colhead{2nd left \tablenotemark{a}} & 
\colhead{1st left} & \colhead{the peak} & 
\colhead{1st right} & \colhead{2nd right} 
}

\startdata
$k$ \tablenotemark{d} & 0.198  & 0.201 & 0.204 & 0.207 & 0.211 \\
$P(k; z=30)$ & 5.442 & 17.75 & 29.66 & 26.11 & 11.53 \\
$P(k; z=0)$  & 2884. & 4664. & 6265. & 5616. & 3632. \\
Smearing \tablenotemark{b}     & 0.96 & 0.48 & 0.38 & 0.39 & 0.57  \\
iVolume \tablenotemark{c}     & 48870 & 51554 & 54090 & 54314 & 56910  
\enddata

\tablenotetext{a}{Relative to the highest point of the peak.}
\tablenotetext{b}{Smearing is $P(k;z=0)D_+^2(z=30)/{[P(k;z=30)
                  D_+^2(z=0)]}$,
                where $D_+(z)$ is the linear growth function.}
\tablenotetext{c}{The number of cells in $k$ space that contribute to
                  the power spectrum. It determines the Poisson noise.}
\tablenotetext{d}{Units: [$k$]=$h {\rm Mpc}^{-1}$; [$P(k)$]=$h^{-3} {\rm Mpc}^3$.}

\end{deluxetable}

   The most conservative estimate of how much does the peak shift is to 
assume that at $z=0$ the maximum of the peak is somewhere between the second 
left ($k=0.198\,h{\rm Mpc}^{-1}$) and second right ($k=0.211\,h{\rm Mpc}^{-1}$) points 
relative to the highest point of the peak in figure\,\ref{fig:peakShift}. 
We also know that the maximum of the input peak is at $k=0.204\,h{\rm Mpc}^{-1}$; 
So if the peak has shifted, the shift is less than 
$(0.211-0.204)/0.204 = 3.2\%$. 

   One reasonable assumption we can make is that there is only one peak.
This narrows down the maximum to somewhere between the first left 
($k=0.201\,h{\rm Mpc}^{-1}$) and first right ($k=0.207\,h{\rm Mpc}^{-1}$) points.  
With the input maximum known, the shift of the peak is less than $1.6\%$.

\begin{figure}[ht]
\begin{center}
\includegraphics[width=3.4in]{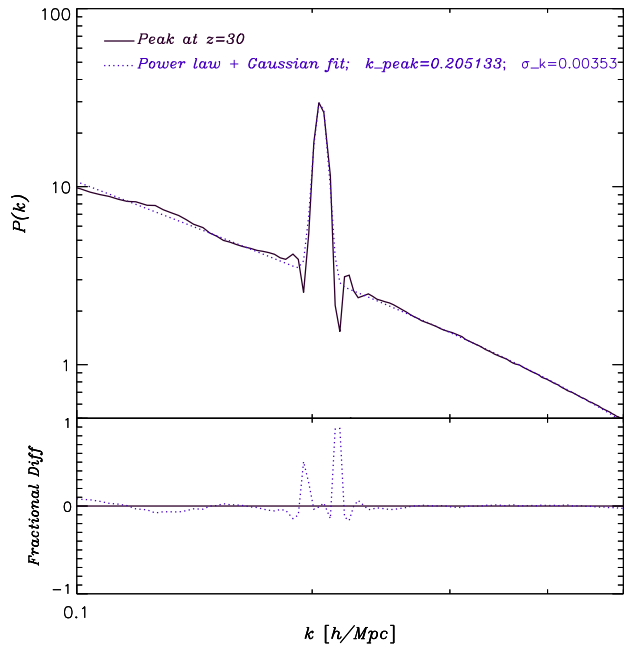}
\end{center}
\caption[Peak at z=30 fitted with Gaussian] {Fitting the $z=30$ power spectrum
                     near the peak with a power law plus a Gaussian.
                     The peak position $k_{\rm peak}$ and width
                     $\sigma_k$ are labeled in the legend.
\label{fig:peakFit_z30}}
\end{figure}

\begin{figure}[ht]
\begin{center}
\includegraphics[width=3.4in]{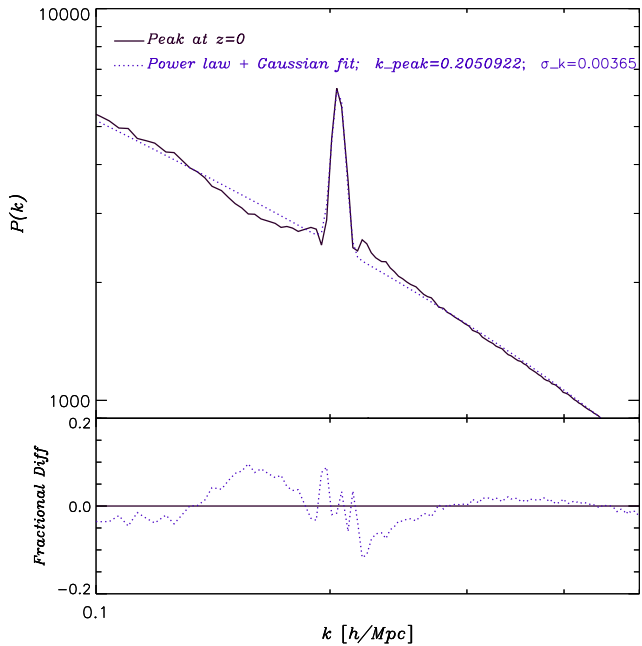}
\end{center}
\caption[Peak at z=0 fitted with Gaussian] {Fitting the $z=0$ power spectrum
                     near the peak with a power law plus a Gaussian.
                     The peak position $k_{\rm peak}$ and width
                     $\sigma_k$ are labeled in the legend.
\label{fig:peakFit_z0}}
\end{figure}

   We can certainly do better than these crude estimates.
One simple method we use is to fit the power spectrum local to the peak 
with a power law plus a Gaussian. The fits for the power spectra at $z=0$
and $z=30$ are shown in figure\,\ref{fig:peakFit_z0} and 
figure\,\ref{fig:peakFit_z30}. The centers of the peak are found to be at 
$0.205092\,h{\rm Mpc}^{-1}$ and $0.205133\,h{\rm Mpc}^{-1}$ respectively. They agree 
to $0.02\%$. This estimation is approaching the theoretical limit
and for any practical purpose this is equivalent to no shift.

   These simple arguments have not taken errors into account. Sample variance
does not apply here because the initial peak and the final peak is from the same
realization. The Poisson errors are on the level of $0.4\%$. This is too small to
change the ordering of the points in the peak, so the first two estimations
are not affected. The last estimation would have an uncertainty of
$\sim 10^{-6}$ which is too small to be relevant.

\subsection{Other effects that could shift the BAO peaks}

   There are effects other than nonlinear collapse could shift the BAO
peaks. The experiment above is dark matter simulation, but the actual
tracer of BAO peaks are galaxies which we know is biased tracer of dark
matter. If galaxy bias
is scale dependent, it could potentially shift the peaks. As shown in
\cite{White05}, and \cite{Smith06}, this is indeed the 
case and the scale dependency should be carefully calibrated.

   Another worry is that BAO peaks sit on top of the smooth underline power 
spectrum whose slope changes due to nonlinear evolution might shift the 
peaks. As pointed out in \cite{Eisenstein_Seo06_1} that this need not be 
the case. The argument is that one can easily make a template of the 
smooth underline power spectrum against which to achieve an unbiased 
measurement.

\subsection{Effects of the peak on nonlinear power spectrum}

   To see the effects of the peak on nonlinear power spectrum,
we compare the nonlinear power spectra evolved
from initial power spectra with and without artificial peak. The smooth part
of these initial power spectra are set to have the same amplitude. Note that
this is different from requiring the same $\sigma_8$ which would allocate more
power to the no peak case on scales outside the peak which
would certainly make things more complicated and less clean. 
The simulation we use for this test is with $N_{\rm p}=256^3$, $N_{\rm g}=512^3$, 
$L_{\rm box}=400 h^{-1}{\rm Mpc}$ and the corresponding
Nyquist frequency is $k_{\rm Ny}=4 h {\rm Mpc}^{-1}$. We start the simulation at $z=30$
and the step size is $da=0.004$. The higher Nyquist frequency is the main 
reason that we use this run instead of the one used for the peak position test. 

\begin{figure}[ht]
\begin{center}
\includegraphics[width=3.4in]{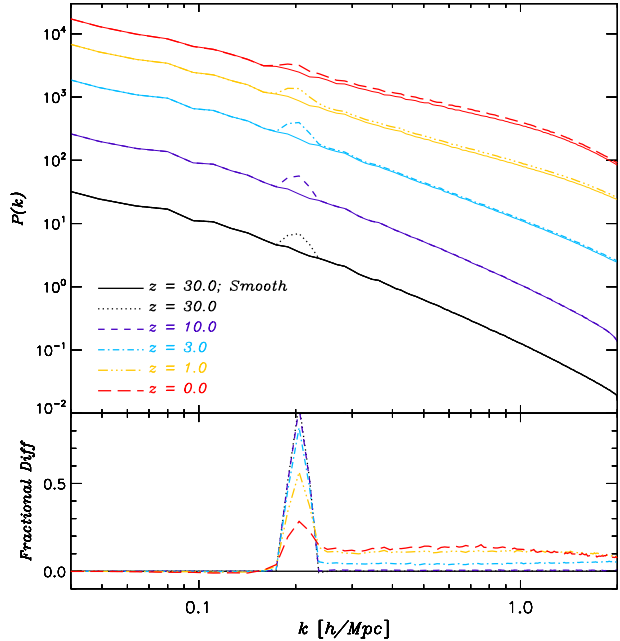}
\end{center}
\caption[Effect of peak to the nonlinear power spectrum] {Comparison of
                 power spectra in simulations with or without a peak in the
                 initial power spectrum. The comparison is done at a few
                 redshifts to show the evolution of the peak and the nonlinear
                 power spectrum. It is clear that \cite{PD96} like mapping
                 between linear and nonlinear $k$ does not exist. Instead,
                 the effect of the peak is close to what halo model
                 would predict.
\label{fig:new_peakPowerGoesTo}}
\end{figure}

   The result is shown in figure\,\ref{fig:new_peakPowerGoesTo}. 
\begin{figure}[ht]
\begin{center}
\includegraphics[width=3.4in]{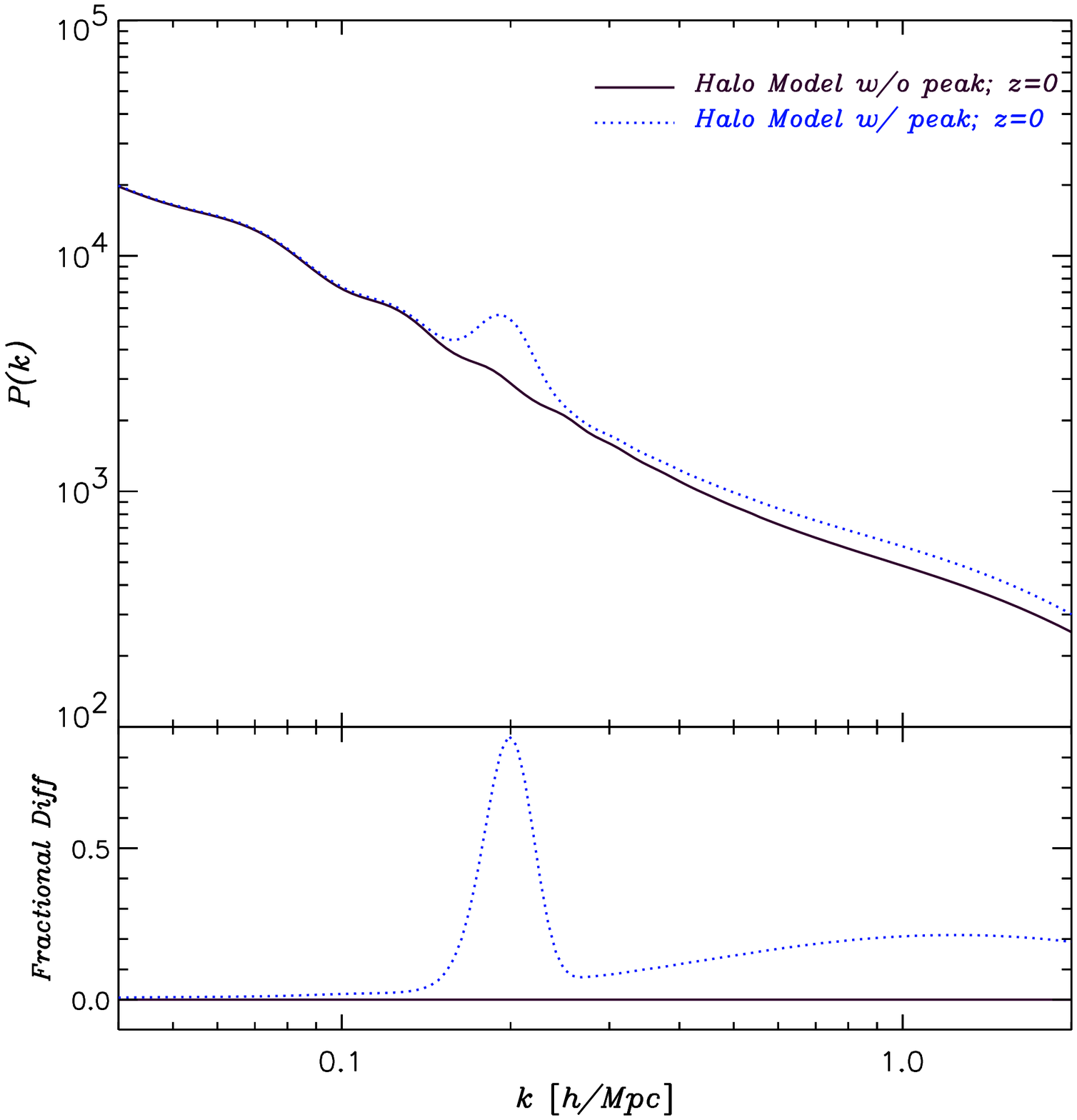}
\end{center}
\caption[Halo model prediction of peak in the power spectrum] {Comparison of
                 power spectra predicted by halo model for the cases with
                 or without a peak in the
                 linear power spectrum. The comparison is done at $z=0$.
                 The effect of the peak on the nonlinear power spectrum
                 is similar to that from simulation
                 (figure\,\ref{fig:new_peakPowerGoesTo}), but different
                 in details.
\label{fig:haloPeak}}
\end{figure}
We see that the peak is washed out as the system evolves and the nonlinear 
power is boosted at all scales smaller than the scale where the peak is 
located. The amount of boost in power shows no scale dependency.

   The mapping between linear and nonlinear scale proposed by \cite{Hamilton91}
and adopted by \cite{PD96} would have predicted that the peak being mapped to a 
nonlinear scale $k_{NL}$ which increases as the nonlinear power
grows. We do not see that kind of mapping here. On the other hand,
the result is similar to what halo model's approach would predict. In the
halo model, the one halo term (see equation\,\ref{eqn:P1h}) is determined by 
the halo profile and the halo mass function $n(m)$. The halo profile has 
nothing to do with whether there is a peak 
in the matter power spectrum or not. 
An extra peak in the linear power spectrum would
increase the halo mass function below certain mass threshold determined by
the position of the peak. As a result, the nonlinear power are boosted on
small scales which is close to what we observe in this test. 
In detail, it does not agree with halo model's prediction perfectly. As 
shown in 
figure\,\ref{fig:haloPeak}, halo model predicts that the fractional boost 
in power due to the existence of a peak has some weak scale dependence 
which is slightly different from what we see in the simulation. 
Since the addition of a spike in the power spectrum only changes the mass
function at the low mass end where $u(k | m)$ is scale independent at
$k = $ a few tenths $h{\rm Mpc}^{-1}$, the one halo term acts
like a shot noise term. The fractional difference of the power spectra 
is scale dependent due to the addition of this short noise like term to 
the power spectrum.
We do not yet understand what accounts for this difference between halo 
model and the simulation.

\section{Linear Growth Function and non-linear Matter Power Spectrum}

   Existing fitting formulas of matter power spectrum, \cite{PD96}, 
\cite{Smith03} or a simple Halo model, have the following assumption:
the nonlinear power spectrum does not depend on 
the time evolution (history) 
of the linear power spectrum; it is determined by the linear power 
spectrum at the same epoch.

   There is a weaker version of the assumption, which states that the
linear power spectrum plus its time evolution determine the nonlinear
power spectrum. A halo model incorporating merger history is a good
example of building fitting formula of nonlinear matter power spectrum
under this weaker assumption.

   In this section, we test both versions of the above assumption using
carefully chosen dark energy models. 

\subsection{Dark energy models construction}

   We choose the set of binned growth suppression $G_i \equiv D_+(a_i)/a_i$ 
and equation of
state of dark energy $w_i$ as parameters to be considered. $G_i$ and $w_i$
share the same binning with $\Delta z = 0.1$. The linear response of $G_i$ 
to the perturbations of $w_i$ can be described by,
\begin{equation}
  \delta G_i = R_{ij} \delta w_j \,.
\label{eqn:Rmat}
\end{equation}
Operationally the perturbation in $w_i$ is achieved by modulating the 
amplitude of the form,
\begin{equation}
  S_w \propto \exp \left [ - {(z-z_i)^{\alpha} \over 
                {2 \sigma_z^2}} \right ] \,,
\end{equation}
where $\alpha = 8$ is used to make the shapelet sharp, and $\sigma_z = 0.05$
is used ($2 \sigma_z = 0.1$ is the binning size). The resulting $w_i$ binnings
are slightly overlapping which does not affect the applications considered in
this work. $G_i$ is simply chosen as the growth suppression at the center
of the redshift bin $z_i$.
   The response matrix $R_{ij}$ is calculated numerically using the formula
in \cite{Hu_Eisenstein99}.

\begin{figure}[ht]
\begin{center}
\includegraphics[width=3.4in]{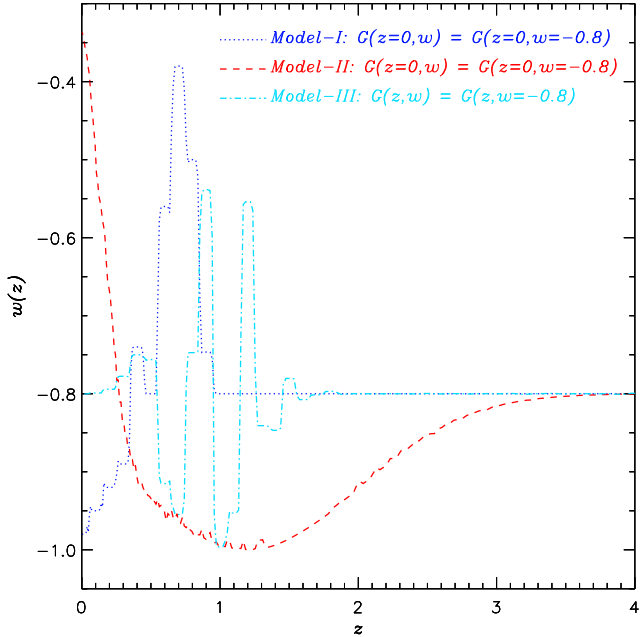}
\end{center}
\caption[w(z) model gives the same growth as w=-0.8 at z=0] { The
              time evolution of the equation of state of dark energy.
              Model-I \& II have the same linear growth functions at
              $z=0$ as that of $w=-0.8$ model but different in the past.
              Their growth function comparisons are shown in 
              figure\,\ref{fig:GdiffHistory}. Model-III has the same
              linear growth function (precise to $0.05\%$ as shown in
              figure\,\ref{fig:GsameHistory}), at all
              redshift, as that of $w=-0.8$ model. 
\label{fig:wSameGz0}}
\end{figure}

\begin{figure}[ht]
\begin{center}
\includegraphics[width=3.4in]{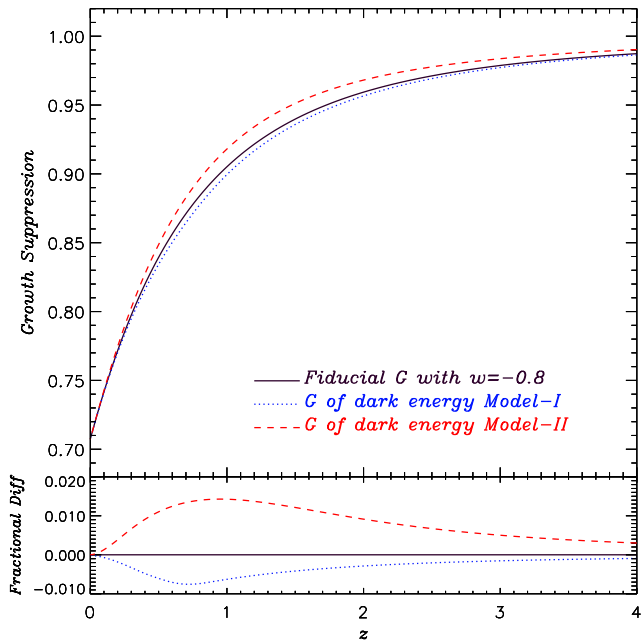}
\end{center}
\caption[Growth with different history] {Comparison of the growth function
             of $w=-0.8$ model with these of the constructed dark energy
             model-I \& II which are shown in figure\,\ref{fig:wSameGz0}.
             These dark energy models have the same growth
             functions at $z=0$, but different time evolutions.
             The fractional differences shown in the lower panel are
             relative to $w=-0.8$ model.
\label{fig:GdiffHistory}}
\end{figure}

   To test the strong assumption, we want to construct a $w$ model that leads
to the same growth factor as the $w=w_{\rm const}$ model at $z=0$ but different growth 
in the past.  
In order to keep the linear growth factor at $z=0$ the same, the $w$ model we
are looking for should have an equation of state greater than $w_{\rm const}$
during some epochs and smaller during some other epochs (to offset the
effect on linear growth when it is greater). So we choose $w_{\rm const}=-0.8$,
instead of -1, to avoid having $w_i < -1$.
There are many ways to achieve our goal. The response matrix $R_{ij}$
is invertible. So one could either specify the linear growth and derive 
the corresponding $w_i$ or the other way around. Here we use the latter.
First we know that the $w$ model being sought after satisfies the
constrain $\delta G_1= 0 = R_{1j} \delta w_j$, where we explicitly label
the redshift bin at $z=0$ as bin number 1. Furthermore, we restrict the
variation of $w$ to below redshift one where the effect of dark energy 
dominates.
Finally we demand that $w(z) > -1$. We then assign values to $\delta w_j$ 
satisfying the above conditions. 
One of our choices of $w_i$ is shown in figure\,\ref{fig:wSameGz0}
(model-I). 
Figure\,\ref{fig:GdiffHistory} shows the comparison of the 
corresponding linear growth suppression to that of $w=-0.8$ model. 
Our construction is successful since the two dark energy models have the
same linear growth at $z=0$ and a maximum deviation from each other
by $0.8\%$ at $z \sim 0.7$.
\begin{figure}[ht]
\begin{center}
\includegraphics[width=3.4in]{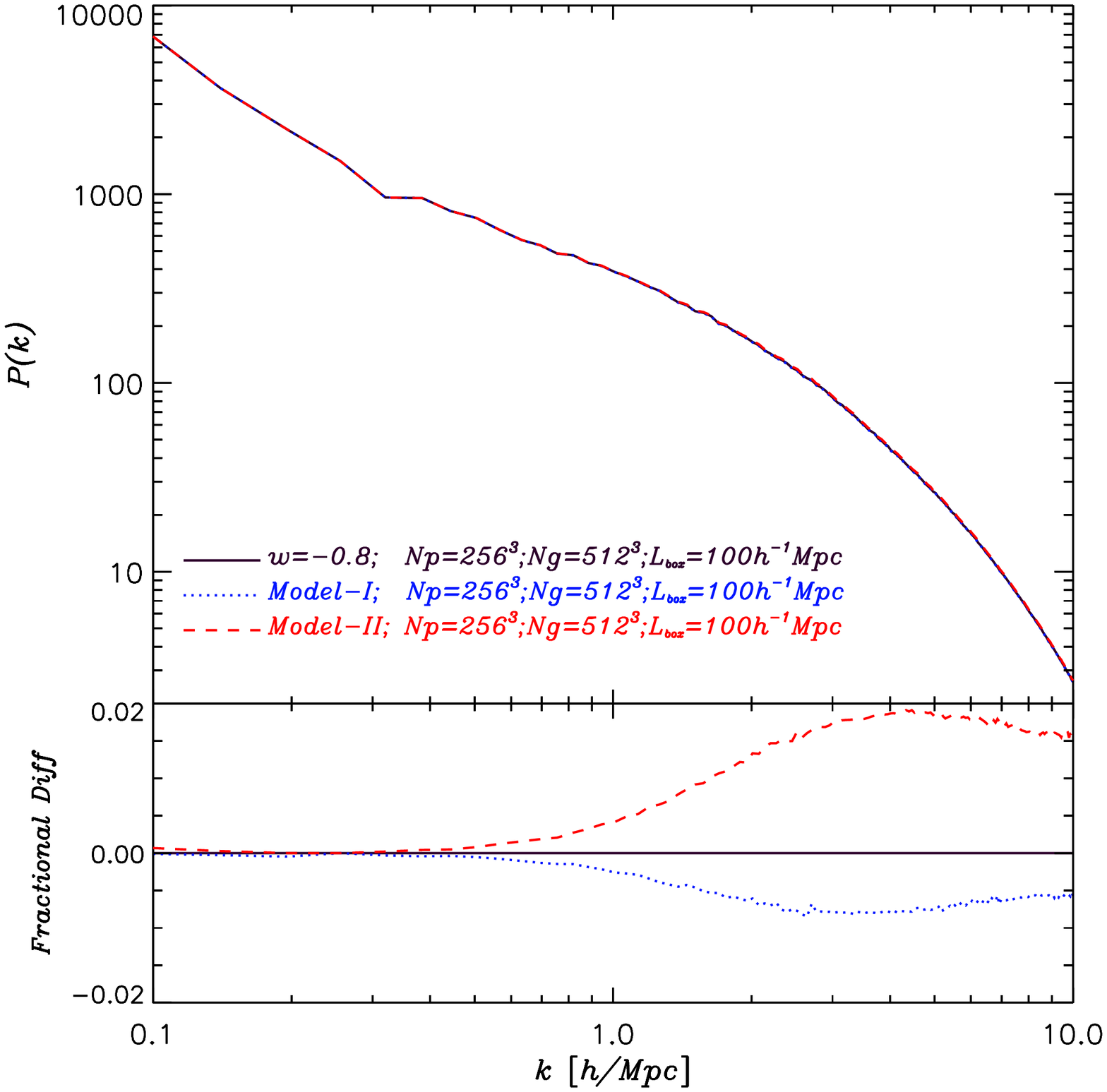}
\end{center}
\caption[Comparison of P(k) with growth history different] {
             Comparison of the matter power
             spectrum at $z=0$ of $w=-0.8$ model with these of the constructed
             dark energy model-I \& II which are shown in 
             figure\,\ref{fig:wSameGz0}.
             These dark energy models have the same growth
             functions at $z=0$ but different time evolutions.
             The fractional differences shown in the lower panel are
             relative to $w=-0.8$ model.
             The simulation box size is $L_{\rm box}=100 h {\rm Mpc}^{-1}$ 
             and the Nyquist frequency is
             $k_{\rm Ny} = 16 h {\rm Mpc}^{-1}$ .
\label{fig:PkGdiffHistoryLbox100}}
\end{figure}

\begin{figure}[ht]
\begin{center}
\includegraphics[width=3.4in]{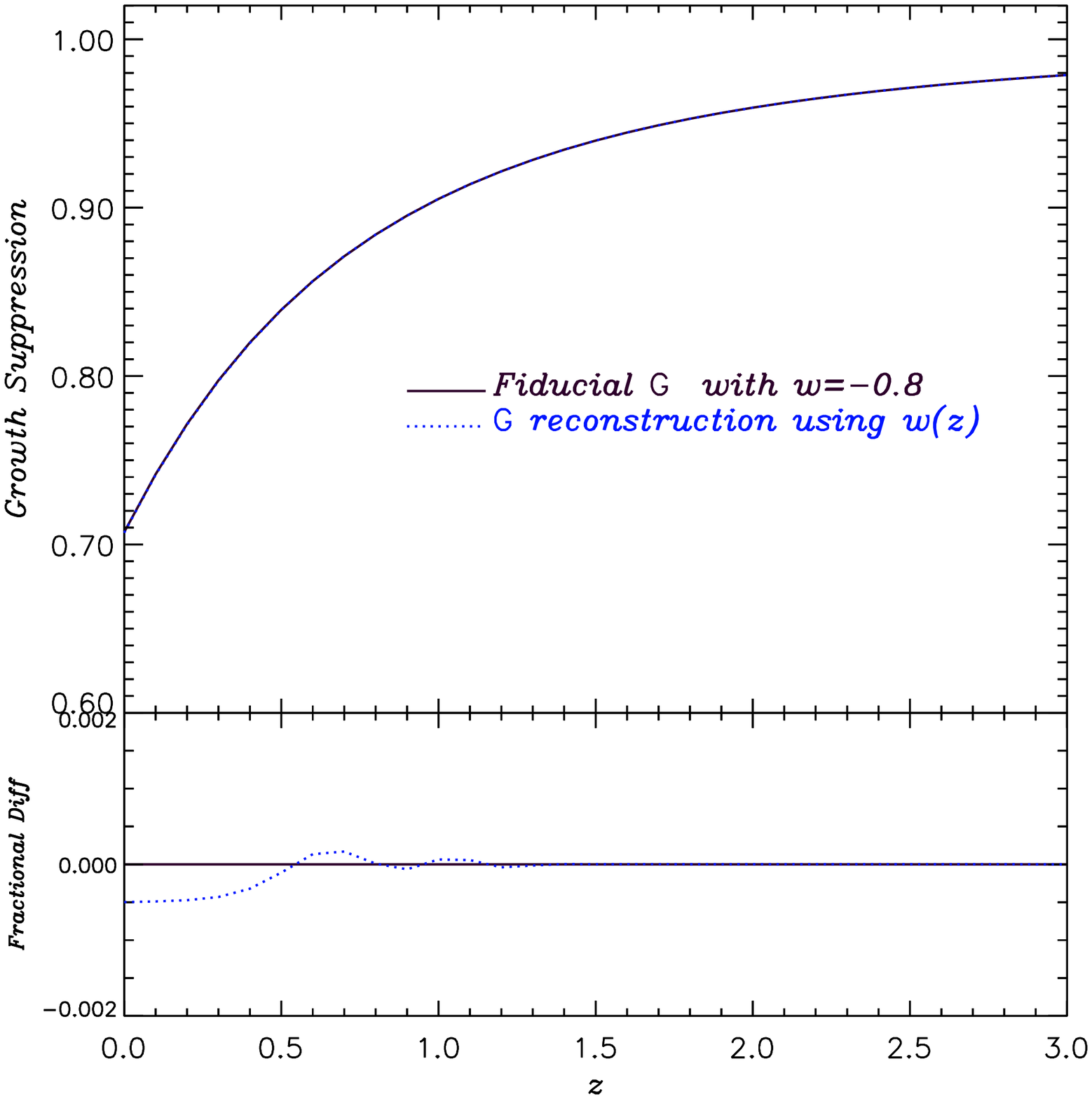}
\end{center}
\caption[Growth with the same history] {Comparison of the growth function
             of $w=-0.8$ model with that of a wiggling
             $w(z)$ model-III which is shown in figure\,\ref{fig:wSameGz0}.
             The linear growth functions of these two dark energy
             models agree to better than $0.05\%$ at all redshift.
\label{fig:GsameHistory}}
\end{figure}

\begin{figure}[ht]
\begin{center}
\includegraphics[width=3.4in]{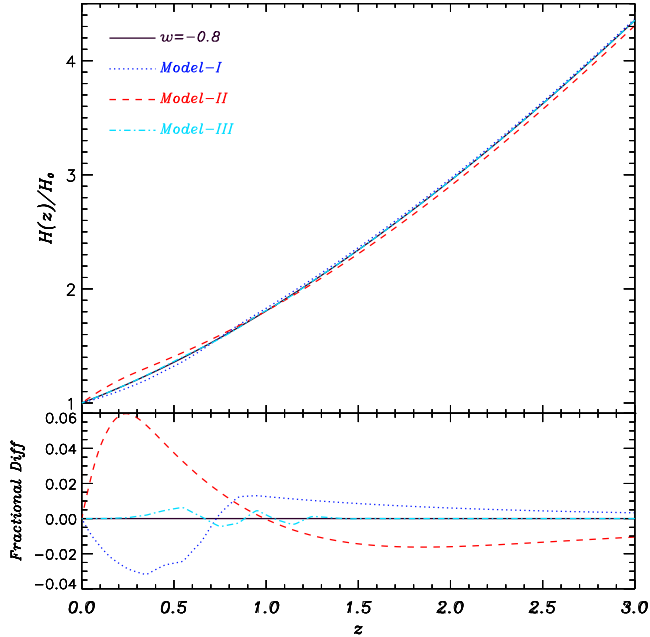}
\end{center}
\caption[Expansion history H/H0] {Comparison of the expansion history
             $H(z)/H_0$ of $w=-0.8$ model with dark energy models whose
             equations of state are shown in
             figure\,\ref{fig:wSameGz0}.
\label{fig:Hz}}
\end{figure}

   To test the weaker version of the assumption, we want to construct a
$w$ model that predicts the same (to certain precision) linear growth 
function as that of $w=-0.8$ model at all redshift.
We apply singular value decomposition (SVD) to the response matrix $R_{ij}$
and select the eigenvectors that have small eigenvalues. These eigenvectors
correspond to the combinations of $\delta w_i$ to which $\delta G_i$ has 
the least response. Among the selected modes we pick the ones that 
involve variations of $w_i$ at low redshift where the effect of dark energy
dominates. Only one mode passed the selection criteria and is shown in
figure\,\ref{fig:wSameGz0} as model-III. The linear growth of the selected $w$ model
agrees with that of $w=-0.8$ model better than $0.05\%$ at all redshift as
shown in figure\,\ref{fig:GsameHistory}. On the other hand, the expansion
histories of these two models differ by $\sim 0.6\%$ as shown in 
figure\,\ref{fig:Hz}.
\begin{figure}[ht]
\begin{center}
\includegraphics[width=3.4in]{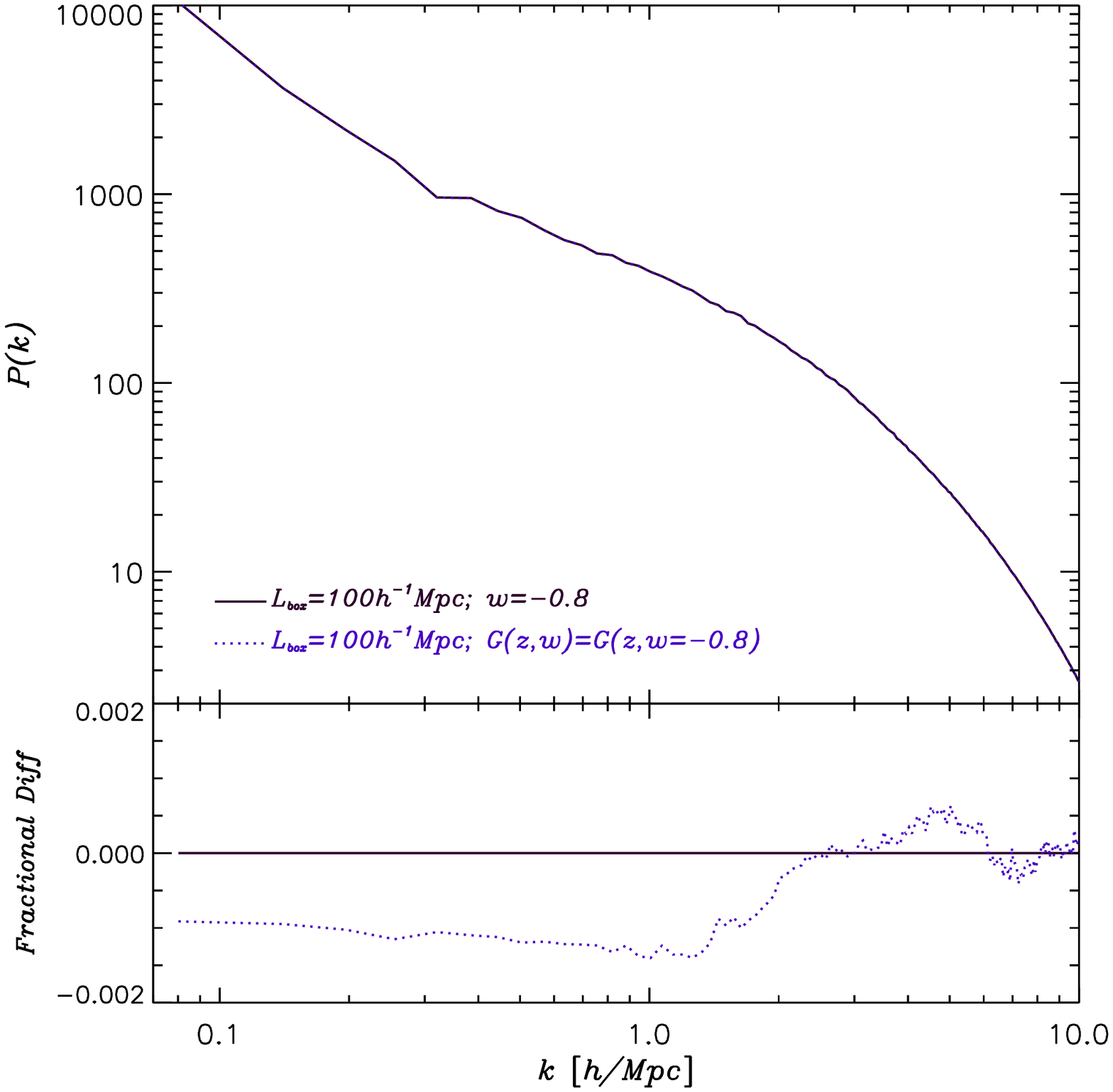}
\end{center}
\caption[Comparison of P(k) with growth history the same] {Comparison
             of the matter power
             spectrum at $z=0$ of $w=-0.8$ model with that of dark energy
             model-III which is shown in figure\,\ref{fig:wSameGz0}.
             The linear growth functions of these two dark energy
             models agree to better than $0.05\%$ at all redshift.
             We conclude that the two $P(k)$ are consistent.
\label{fig:PkGsameHistory}}
\end{figure}

\subsection{History does matter}

   Figure\,\ref{fig:PkGdiffHistoryLbox100} and \ref{fig:PkGsameHistory} 
summarize the results. As shown in figure\,\ref{fig:PkGdiffHistoryLbox100}, 
the nonlinear
matter power spectra at $z=0$ are different between $w=-0.8$ model and the 
constructed dark energy model-I (figure\,\ref{fig:wSameGz0}) which has the 
same linear 
growth at $z=0$ but different during the past. So the nonlinear power 
spectrum does care about the time evolution of the linear power spectrum, 
contrary to what existing fitting formulas assume. Our interpretation is
the following.
The difference in the linear power spectrum during the past causes some
differences in the nonlinear power spectrum which would stay there unless
some rather delicate adjustments of the history of the linear power 
spectrum is done to offset the effect. An arbitrary history of the linear
power spectrum does not have the ability to undo the difference in the 
nonlinear power spectrum. One possibility is that different linear growth
histories would produce different halo merger histories on which the halo 
concentration depends \citep{Weshsler05}. So the nonlinear power spectrum
which mainly comes from the one halo term would be different. It is hard
to imagine that simply matching the linear growth at $z=0$ would erase
the difference. So in general the difference would be there as shown in our
test case.

   The effect of linear growth history on nonlinear power spectrum is by
no means limited to the $\sim 0.8\%$ level shown in our test case. The
nonsmooth nature of $w(z)$ also would not make our argument any weaker. 
By a little bit of more work, we come up with another example that has much 
smoother $w(z)$ curve (figure\,\ref{fig:wSameGz0} model-II) and 
matching linear growth at $z=0$ but maximum difference of
$\sim 1.5\%$ in the past (figure\,\ref{fig:GdiffHistory} dash line). 
As shown in figure\,\ref{fig:PkGdiffHistoryLbox100} (dash line), the resulting 
nonlinear power spectra have maximum difference at $\sim 2\%$ level.
From these examples, we see that the difference of the nonlinear power
spectrum due to different linear growth history is at the level of the 
maximum deviation of the corresponding linear power spectra in the past.

   Our test of the weaker assumption validates it as shown in
figure\,\ref{fig:PkGsameHistory}. The nonlinear power spectra of $w=-0.8$ 
model and the constructed dark energy model-III (figure\,\ref{fig:wSameGz0}) are 
consistent with being essentially the same given the small difference of 
the linear power spectra. This is not a proof that the history of the
linear power spectrum fully determines the nonlinear power spectrum,
but it hints that it is possibly the case. 
Note that although there is basically no hope of distinguishing the two
apparently very different dark energy models using their growth functions,
their expansion histories differ by $\sim 0.6\%$ (see figure\,\ref{fig:Hz})
which is close to the level that BAO could potentially have 
some handle on.

\section{Matter Power Spectrum and the Equation of State of Dark Energy}

   With the ambitious observational efforts focused on measuring the
equation of state of dark energy to high precision 
(DES,SNAP,LSST,PanSTARRS), 
theoretical parameter forecasts will need to be done more carefully than before.
In particular, we have to have a better understanding of the nonlinear power 
spectrum which is very important for cosmic shear and any other probes that
utilize the information of growth of density perturbations. Since numerical
simulations are too expensive to run for every application, fitting formulas
of the matter power spectrum are widely used.
The ones provided by
\cite{PD96} and \cite{Smith03} are calibrated using $\Lambda$CDM models. 
\cite{McDonald05} provided a correction factor to the above mentioned 
fitting formulas for cosmology with constant $w$ \citep[see also][]{Ma99}.
For two parameter dark energy models parameterized by
$w_0$ and $w_a$, \cite{Linder_White05} gives a simple procedure to calculate
the nonlinear matter power spectrum with a few percent accuracy.
While the method provided by \cite{Linder_White05} has not yet been widely
adopted,
a common practice is to apply \cite{PD96} and \cite{Smith03} fitting formulas
regardless of the dark energy models.
In this work we evaluate how well this common practice works.

    Two things could go wrong if we used an inaccurate fitting formula.
One is that the fitting formula predicts the wrong amplitude of the matter
power spectrum. This is the part we study in this work. Another effect of
incorrect fitting formula is that the predicted model parameter degeneracy
directions and/or extent of the directions are off. Although not covered 
in this study, the latter effect is very important when the parameters are
highly degenerate.
During our study of the effect of photo-z on weak lensing tomography 
\citep{MaHu05}, we noticed that with no tomography binning the extent of the
degeneracy among dark energy parameters could differ by a factor of 4
depending on whether the fitting forms of \cite{PD96} or halo model 
is used to calculate the matter power spectrum.

    For this study we further restrict ourself to two parameter dark energy 
models which are parametrized by $w_0$ and $w_a$.

\subsection{The findings}

\begin{figure}[ht]
\begin{center}
\includegraphics[width=3.4in]{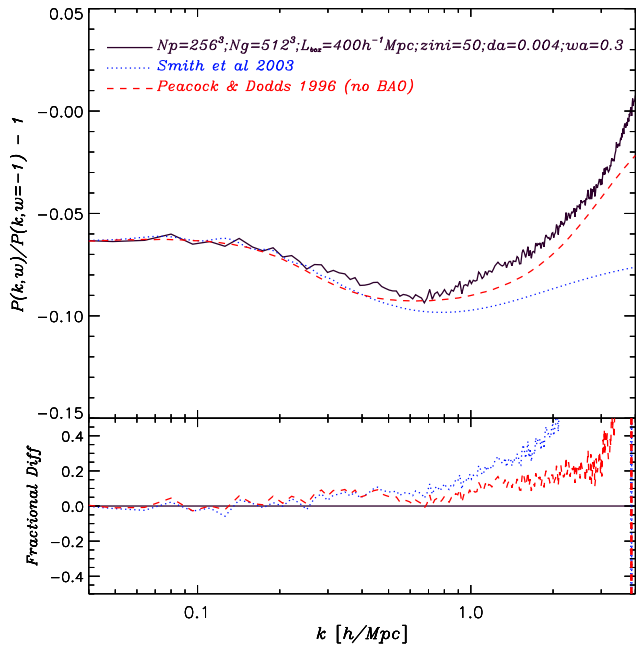}
\end{center}
\caption[P(k) with wa=0.3 Lbox=400 zini=50] {
                    Performance of the fitting formulas when applied to
                    non-$\Lambda$CDM cosmology. In this particular case,
                    $w_0=-1.0$ and $w_a=0.3$.  The simulation
                    box is 400 $h^{-1} {\rm Mpc}$ and the corresponding
                    Nyquist frequency is $k_{\rm Ny} = 4 \,h {\rm Mpc}^{-1}$.
                    The initial power spectrum without BAO peaks is used for
                    the Peacock \& Dodds fitting formula.
\label{fig:wa0.3Lbox400zini50}}
\end{figure}

\begin{figure}[ht]
\begin{center}
\includegraphics[width=3.4in]{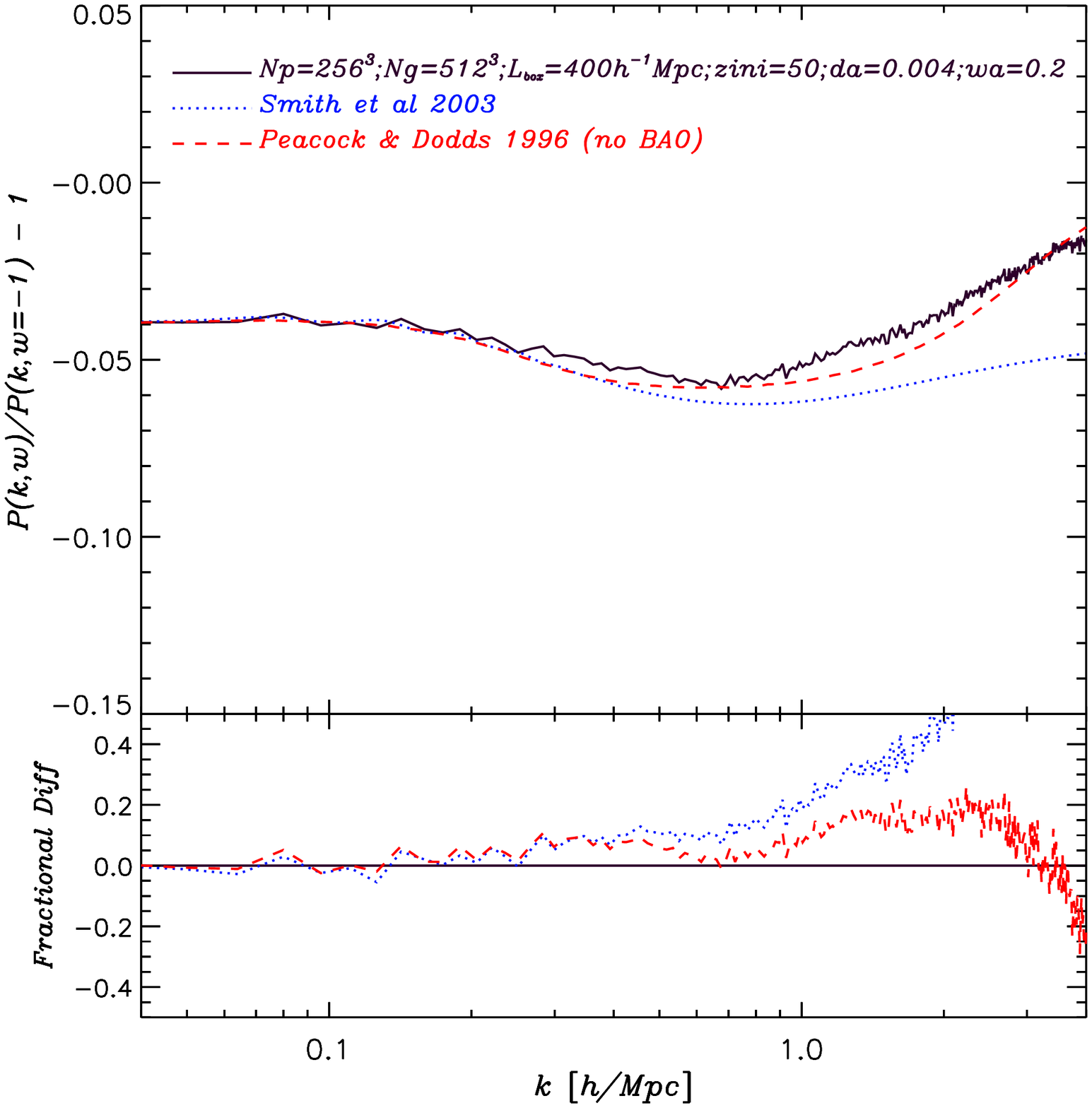}
\end{center}
\caption[P(k) with wa=0.2 Lbox=400 zini=50] {
                    Performance of the fitting formulas when applied to
                    non-$\Lambda$CDM cosmology. In this particular case,
                    $w_0=-1.0$ and $w_a=0.2$.  The simulation
                    box is 400 $h^{-1} {\rm Mpc}$ and the corresponding
                    Nyquist frequency is $k_{\rm Ny} = 4 \,h {\rm Mpc}^{-1}$.
                    The initial power spectrum without BAO peaks is used for
                    the Peacock \& Dodds fitting formula.
\label{fig:wa0.2Lbox400zini50}}
\end{figure}

\begin{figure}[ht]
\begin{center}
\includegraphics[width=3.4in]{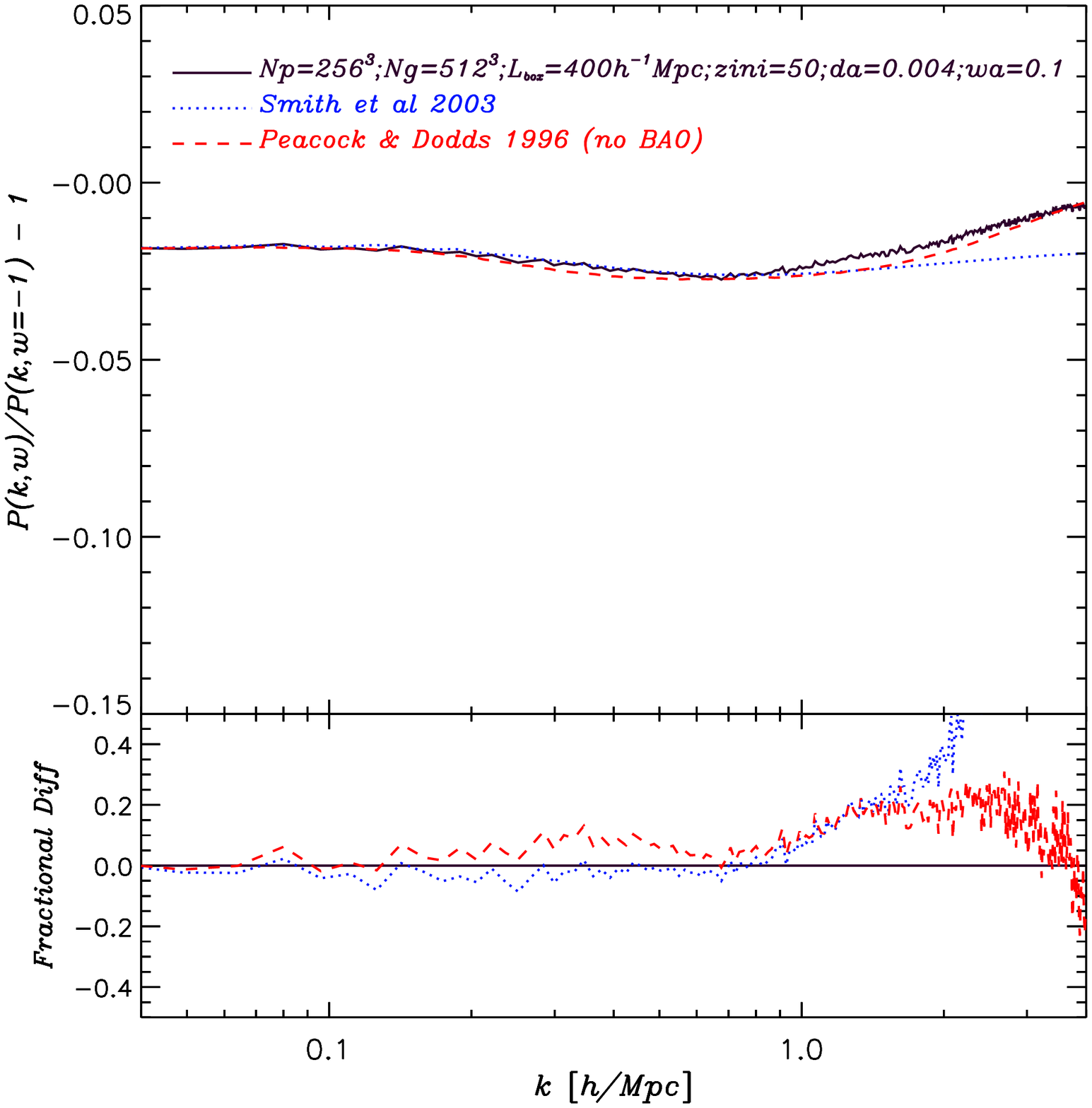}
\end{center}
\caption[P(k) with wa=0.1 Lbox=400 zini=50] {
                    Performance of the fitting formulas when applied to
                    non-$\Lambda$CDM cosmology. In this particular case,
                    $w_0=-1.0$ and $w_a=0.1$.  The simulation
                    box is 400 $h^{-1} {\rm Mpc}$ and the corresponding
                    Nyquist frequency is $k_{\rm Ny} = 4 \,h {\rm Mpc}^{-1}$.
                    The initial power spectrum without BAO peaks is used for
                    the Peacock \& Dodds fitting formula.
\label{fig:wa0.1Lbox400zini50}}
\end{figure}

    The quantity we use to evaluate the performance of the fitting formulas
applied to $w_a \neq 0$ dark energy models is,
\begin{equation}
  Q \equiv {P(k, w) \over {P(k, w=-1)}} -1 ,
\label{eqn:Qfact}
\end{equation}
which is the fractional difference of
the matter power spectrum between $w_a$ model, $P(k, w)$, and the $\Lambda$CDM 
model, ${P(k, w=-1)}$. The fractional difference of the power spectrum in
terms of $Q$ is $\Delta Q / (Q+1)$ and that of the derivative of the power
spectrum is  $\Delta Q / Q \equiv Q(fitting\,\,formula)/Q(simulation)-1$.
Figures\,\ref{fig:wa0.3Lbox400zini50}-\ref{fig:wa0.1Lbox400zini50} show
the comparison of this quantify $Q$ from the simulation with 
those from the fitting formulas of \cite{PD96} and \cite{Smith03} 
for $w_a = 0.3$, $0.2$ and $0.1$ respectively.

    First of all, the fitting formulas are doing a good job describing
$w_a$ models. The fractional difference of the power spectrum is
$\Delta Q / (Q+1) \sim 1\%$ and that of the derivative
is $\Delta Q / Q \sim 10\%$ for $k < 1.0 \,h {\rm Mpc}^{-1}$. 
In the context of
Fisher matrix analysis, the Fisher matrix elements would be $20\%$
off at most. Again, just to remind ourself that we are only considering
the effect of the amplitude of the power spectrum but not that of the
shape of the power spectrum which is closely tied to the degeneracy direction
of the model parameters.

\begin{figure}[ht]
\begin{center}
\includegraphics[width=3.4in]{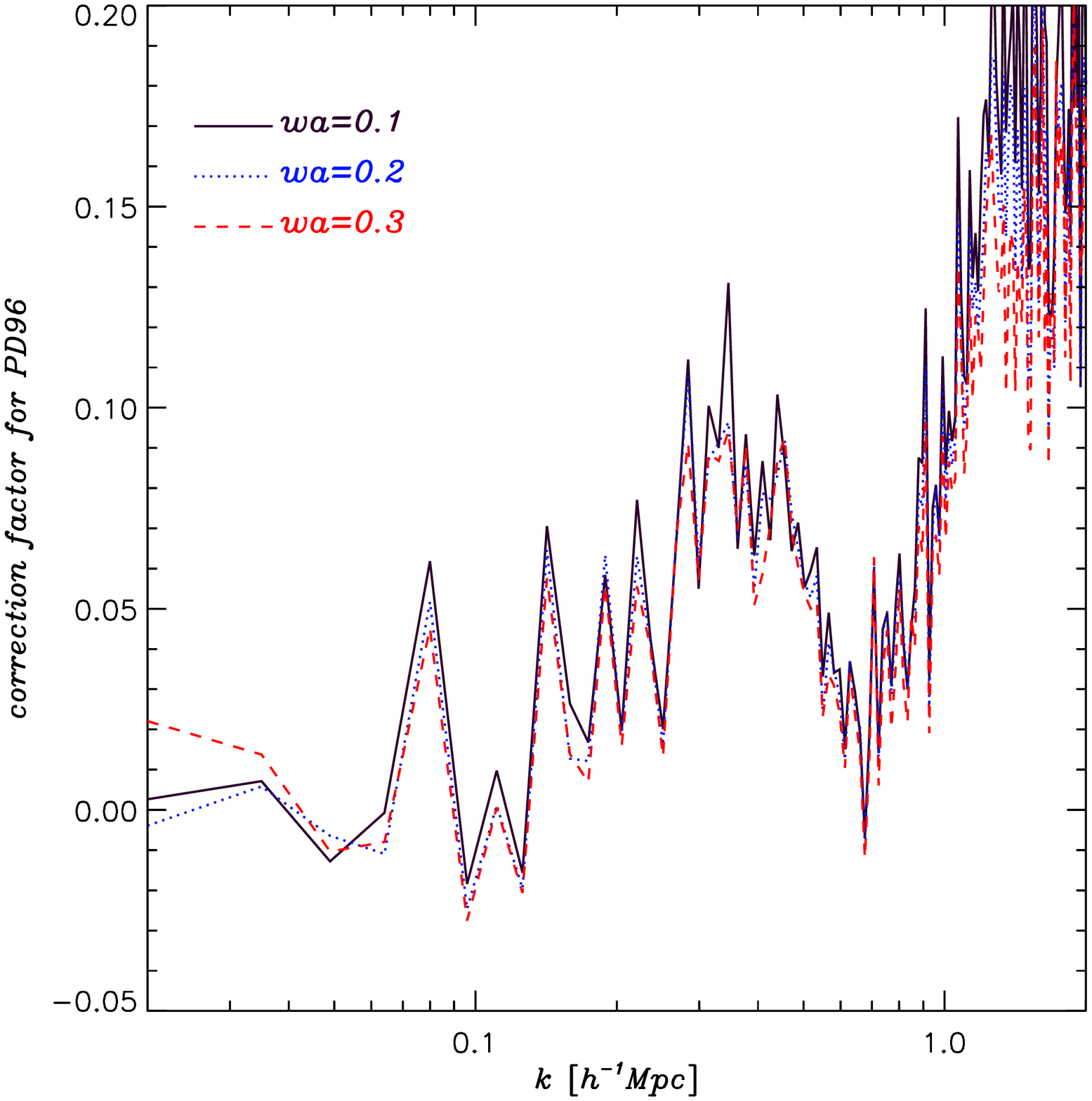}
\end{center}
\caption[Correction factor for PD96 (z=0.0)] {The fractional differences
                     of $Q$ (defined in equation\,\ref{eqn:Qfact}),
                     $Q(fitting\,\,formula)/Q(simulation)-1$,
                     for different $w_a$. The power spectrum is evaluated
                     at $z=0$.
\label{fig:Qfactz0}}
\end{figure}

\begin{figure}[ht]
\begin{center}
\includegraphics[width=3.4in]{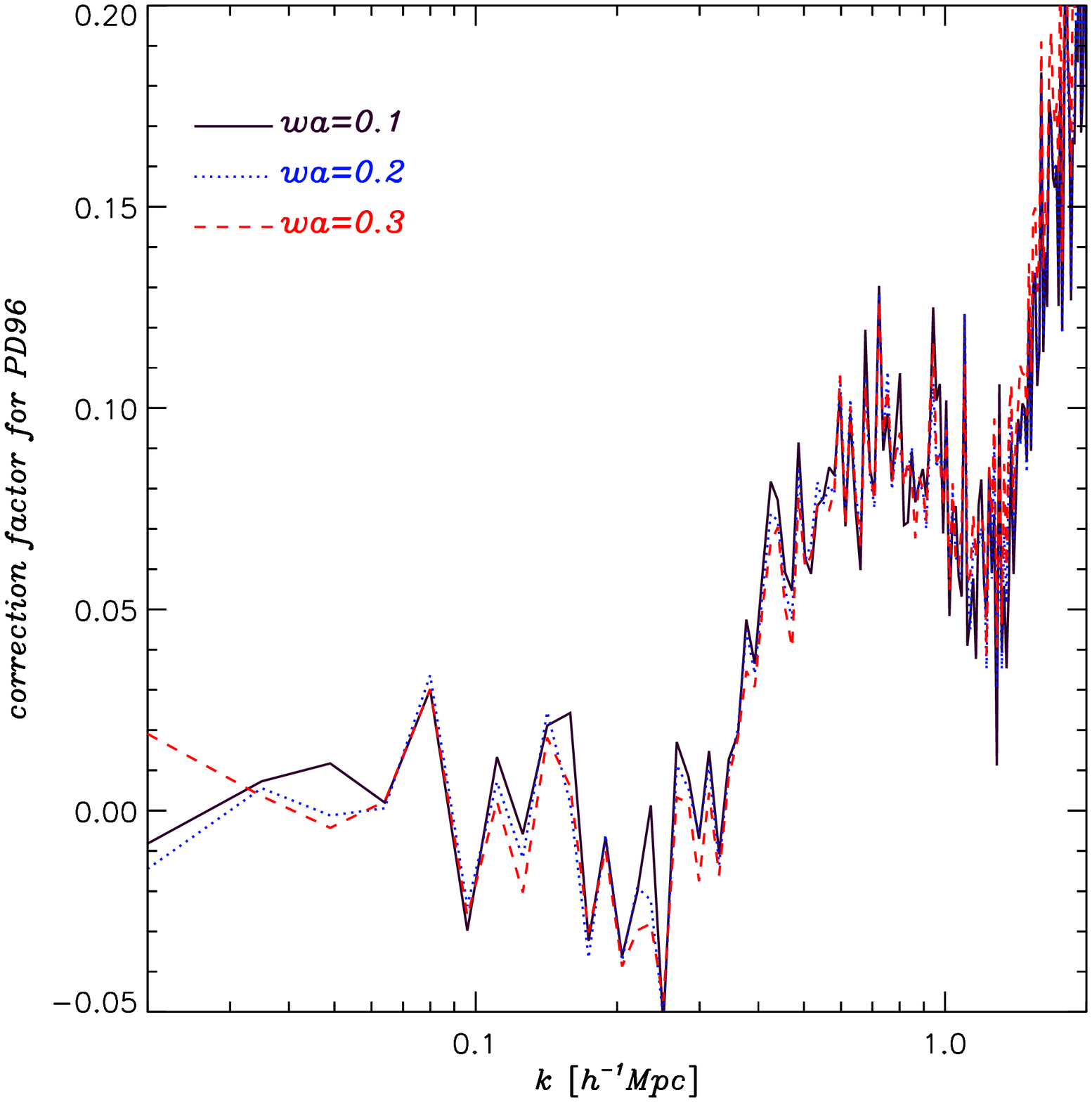}
\end{center}
\caption[Correction factor for PD96 (z=0.62)] {The fractional differences
                     of $Q$ (defined in equation\,\ref{eqn:Qfact}),
                     $Q(fitting\,\,formula)/Q(simulation)-1$,
                     for different $w_a$. The power spectrum is evaluated
                     at $z=0.62$.
\label{fig:Qfactz0.62}}
\end{figure}

    The most surprising result we find is that the fractional differences 
in $Q$ from \cite{PD96} fitting formula are independent of $w_a$, 
as shown in figure\,\ref{fig:Qfactz0}.
To turn this fact into a fitting formula of the matter power
spectrum for $w_a$ models, the time evolution of this
potential correction factor to \cite{PD96} fitting formula should also be
independent of $w_a$. 
Figure\,\ref{fig:Qfactz0.62} shows this correction factor at $z=0.62$. 
We see that the peak of the  correction shifts to smaller scale 
(higher $k$) as one goes to higher redshift and, amazingly, it shows no 
dependency on $w_a$.
Although such regular behavior seems always have a good reason behind it, 
at the moment, we do not have a physical explanation. 

    It is very tempting to use such a fact to come up with a correction
factor to the fitting formula of \cite{PD96}.
The hump at the nonlinear scale in figure\,\ref{fig:Qfactz0} and the ones
at high redshift can be fit very well using a Gaussian in $k$ space. However,
as we have pointed out throughout this work, the physical arguments
supporting Peacock \& Dodds' fitting formula do not seem to hold: 
the linear and nonlinear scale mapping does not hold; the assumption that
the history of the linear growth does not affect the nonlinear power
spectrum is also being shown to be incorrect. Adding more corrections to
Peacock \& Dodds' fitting formula would only turn it into another "theory
of epicycles". We do not think this is the right approach to build a good
fitting formula. As a short term solution, this may be worth doing. We leave
the readers to make the judgment.

\section{Discussion and Conclusions} \label{sec-conclude}

   We modify the public PM code developed by Anatoly Klypin and Jon Holtzman
to simulate cosmologies with arbitrary initial matter power spectra and 
the equation of state of dark energy. Using N-body simulation, we test
various aspects of the matter power spectrum, including
whether nonlinear evolution would shift the BAO peaks, 
the assumptions that go into
constructing fitting formulas of the nonlinear power spectrum, 
and the precision of the existing fitting formulas when applied to dark 
energy models parametrized by $w_0$ and $w_a$.

   Any shift of the BAO peaks in the matter power spectrum would bias the
cosmological distance inferred hence the cosmological parameters derived.
To test whether nonlinear evolution would shift BAO peaks, we implement
an artificial sharp peak in the initial power spectrum and evolve it
using the PM code. We find that the position of the peak is not 
shifted by nonlinear evolution. 
An upper limit of the shift at the level of $0.02\%$ is achieved by fitting
the power spectrum local to the peak using a power law plus a Gaussian.
This implies that, for any practical purpose, the baryon acoustic 
oscillation peaks in the matter power spectrum are not shifted by 
nonlinear evolution. There are other effects such as galaxy bias that could
potentially shift the peaks. These should be carefully calibrated as well.

   We also find that the existence of a peak in the linear power
spectrum would boost the nonlinear power at all scales evenly. This is 
contrary to what HKLM scaling relation predicts, but roughly consistent 
with that of halo model. The scale dependence is slightly different
from halo model prediction in detail. Further study is required to understand the
cause of the difference.

   All existing fitting formulas of the nonlinear power spectrum assume that
the linear power spectrum uniquely determines the nonlinear one, regardless
of the linear growth history.  To test this assumption, 
we construct two dark energy models with the same linear power spectra
today but different linear growth histories and compare their nonlinear power
spectra from the simulation. We find  that the resulting nonlinear
power spectra differ at the level of the maximum deviation of the 
corresponding linear power spectra in the past. 
Similarly, two constructed dark energy
models with the same growth histories result in consistent nonlinear power
spectra. This is consistent with but not a proof of the conventional 
wisdom that together, the linear power spectrum and the
linear growth history uniquely determine the nonlinear power spectrum.

    Next generation large-scale structure surveys need better fitting formulas
than what are available now. With the high accuracy required, new
fitting formulas should be based on solid foundations. Our results suggest 
that we should abandon HKLM scaling relation, keep halo model and further 
include linear growth history to build the next generation fitting formulas 
of nonlinear power spectrum. Note that our study only includes dark matter. To
build fitting formulas with percent level accuracy, one has to include the
effect of baryons \citep{White04,Zhan04,Jing06} and substructure \citep{Hagan05}.

   For simple dark energy models parametrized by $w_0$ and $w_a$, the existing
nonlinear power spectrum
fitting formulas, which are calibrated for $\Lambda$CDM model, work reasonably
well. The corrections needed are at percent level for the power spectrum and
$10\%$ level for the derivative of the power spectrum. 
We find surprisingly that, for \cite{PD96} fitting formula, the corrections 
needed to the derivative of the power spectrum are independent 
of $w_a$ but changing with redshift. As a short term solution, a fitting form 
could be developed for $w_0$, $w_a$ models based on this fact.

\acknowledgments{
                 
    We thank Wayne Hu, Andrey Kravtsov, Scott Dodelson, Josh Frieman, 
Eduardo Rozo and Doug Rudd for useful discussions. We are thankful for 
Anatoly Klypin and Jon Holtzman for making their PM code public.
ZM is supported by David and Lucille Packard Foundation. }

\bibliographystyle{apj}

\end{document}